\documentclass[12pt]{article}
\usepackage{dcolumn}
\usepackage{url}
\newenvironment{wileykeywords}{\textsf{Keywords:}\hspace{\stretch{1}}}{\hspace{\stretch{1}}\rule{1ex}{1ex}}

\usepackage{amsmath,amssymb}
\usepackage{graphicx}
\usepackage{color}
\usepackage{dcolumn}
\usepackage{bm}
\usepackage[numbers,super,comma,sort&compress]{natbib}

\definecolor{background-color}{gray}{0.98}
\newcommand{\singlespace}{\tiny\renewcommand{\baselinestretch}{1.0}\normalsize}

\title{Density-Functional Theory (DFT) and Time-Dependent DFT Study of 
the Chemical and Physical Origins of Key Photoproperties of 
End-Group Derivatives of the Nonfullerene Bulk Heterojunction
Organic Solar Cell Acceptor Molecule IDIC}
\author{Taouali W\thanks{Laboratoire de Recherche (LR18ES19),
Synth\`ese Asym\'etrique et Ing\'enierie Mol\'eculaire
de Mat\'eriaux Organiques pour l'\'Electroniques
Organiques, Facult\'e des Sciences de Monastir,
Universit\'e de Monastir, 5000 Monastir, TUNISIA}, Alimi K\footnotemark[1], Nangraj A.S. \thanks{State Key Laboratory of Microbial Metabolism, Shanghai Jiao Tong University, Shanghai,
CHINA}, Casida M.E.\thanks{Laboratoire de Spectrom\'etrie, Interactions et Chimie th\'eorique
(SITh),
D\'epartement de Chimie Mol\'eculaire (DCM, UMR CNRS/UGA 5250),
Institut de Chimie Mol\'eculaire de Grenoble (ICMG, FR2607),
Universit\'e Grenoble Alpes (UGA)
301 rue de la Chimie, BP 53, F-38041 Grenoble Cedex 9, FRANCE}}
\date{}
\begin{document}
\maketitle







  
\begin{abstract}

As emphasized in a recent review article [{\em Chem.\ Rev.} {\bf 122}, 14180 (2022)], organic solar cell (OSC) photoconversion efficiency has been rapidly evolving
with results increasingly comparable to those of traditional inorganic solar
cells. Historically, OSC performance improvement focused first on the morphology of
P3HT:PC$_{61}$BM solar cells then went through different stages to shift lately interest towards nonfullerene acceptors (NFAs) as a replacement of PC$_{61}$BM  acceptor (ACC) molecule. Here, we use density-functional theory (DFT) and time-dependent (TD) DFT to investigate four novel NFAs of A-D-A (acceptor-donor-acceptor) form derived
from the recently synthesized IDIC-4Cl [{\em Dyes and Pigments} {\bf 166}, 
196 (2019)]. Our
level of theory is carefully evaluted for IDIC-4Cl and then applied to the four novel
NFAs in order to understand how chemical modifications lead to physical changes in
cyclic voltammetry (CV) frontier molecular orbital (FMO) energies and absorption
spectra in solution.Finally we design and apply a new type of Scharber plot for
NFAs based upon some simple but we think reasonable assumptions. Unlike the
original Scharber plots where a larger DON band gap favors a larger PCE, our
modified Scharber plot reflects the fact that a smaller ACC band gap may favor
PCE by filling in gaps in the DON acceptor spectrum. We predict that only the
candidate molecule with the {\em least} good acceptor A, with the {\em highest} frontier
molecular orbital energies, and {\em one of the larger} CV lowest unoccupied molecular
orbital (LUMO) $-$ highest unoccupied molecular orbital (HOMO) gaps, will yield
a PM6:ACC PCE exceeding that of the parent IDIC-4Cl ACC. This candidate
also shows the largest oscillator strength for the primary $^1$(HOMO,LUMO) charge-
transfer transition and the largest degree of delocalization of charge
transfer of any of the ACC molecules investigated here.

\end{abstract}

\vspace{0.5cm}

\noindent
\begin{wileykeywords}
density-functional theory, time-dependent density-functional theory, transition density matrix , organic solar cell, nonfullerene acceptors.
\end{wileykeywords}

\begin{figure}[h]
\centering
\colorbox{background-color}{
\fbox{
\begin{minipage}{1.0\textwidth}
\includegraphics[width=50mm,height=50mm]{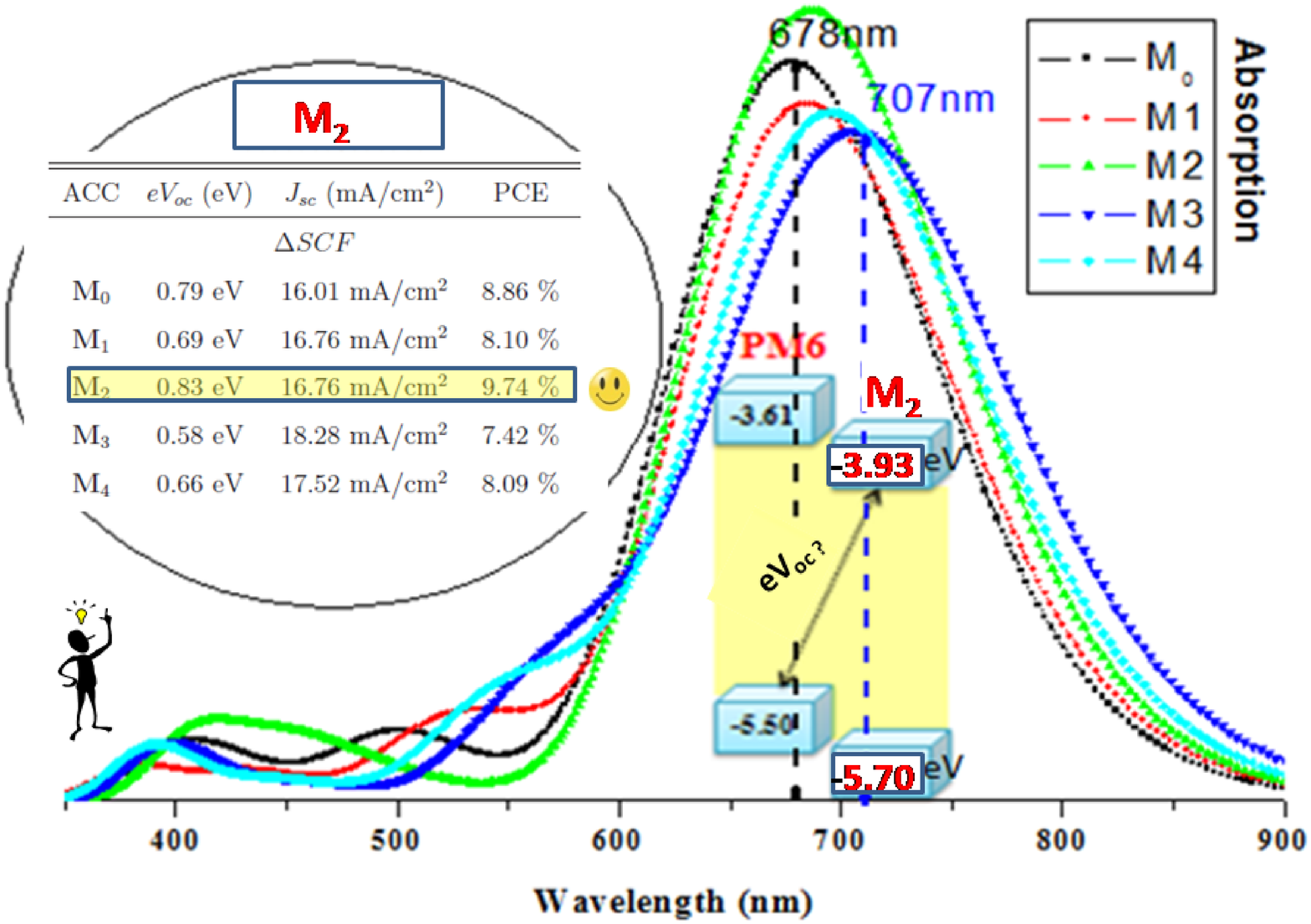} 
\\
(High performance non -fullerene acceptors (NFAs) based on the recently synthesized IDIC-4CL, that can be used to build bulk heterojunction photovoltaic cells, with enhanced open circuit voltage (Voc).)
\end{minipage}
}}
\end{figure}

  \makeatletter
  \renewcommand\@biblabel[1]{#1.}
  \makeatother

\bibliographystyle{jcc}

\renewcommand{\baselinestretch}{1.5}
\normalsize

\clearpage
\section{Introduction}
\label{sec:intro}

\begin{figure}
\begin{center}
\includegraphics[width=0.9\textwidth]{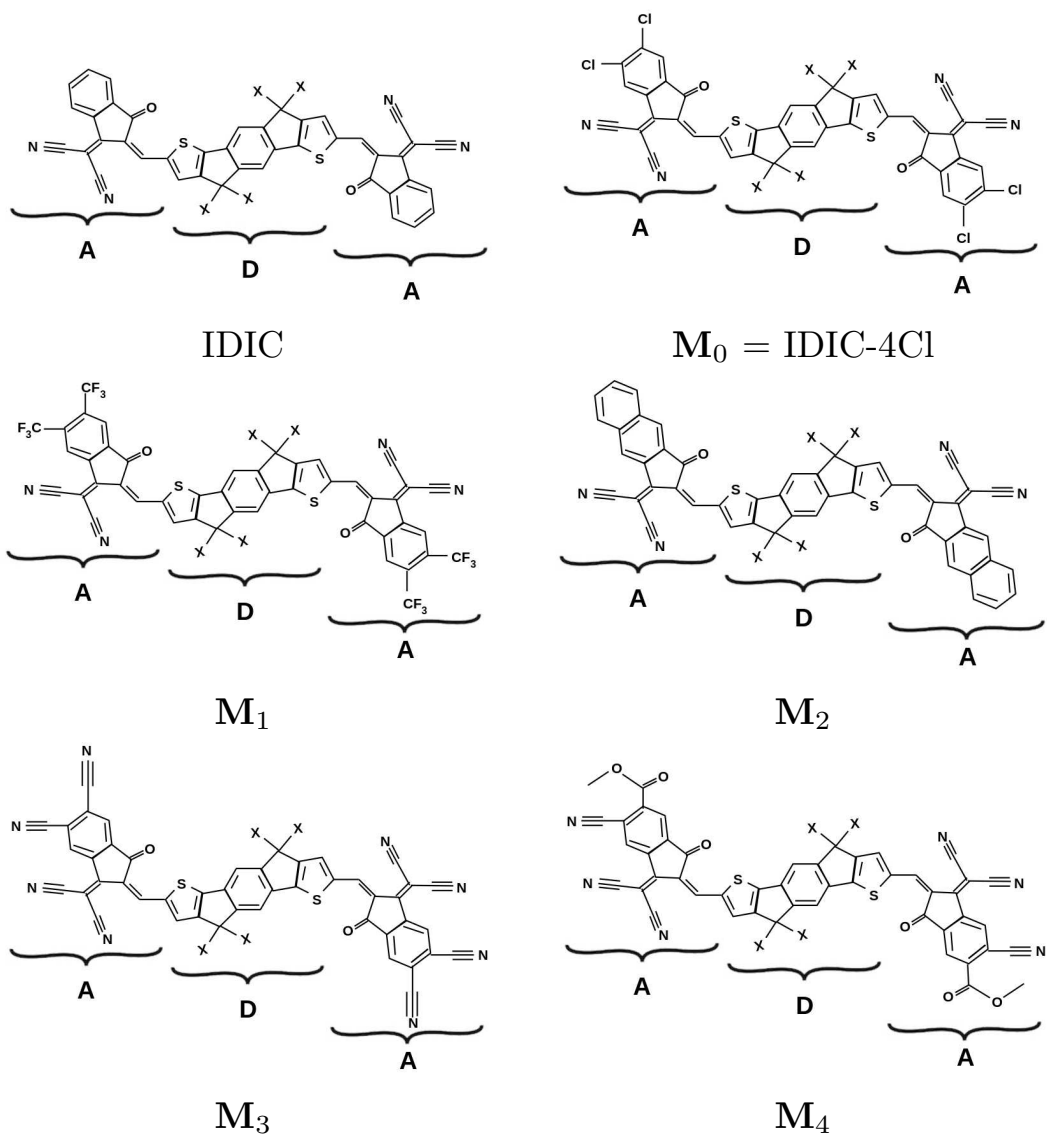}

\end{center}
\caption{
The chemical structures of IDIC and the non-fullerene ACC (NFAs)
treated in this article.  In our work X is a methyl group, but in 
the literature X is typically an $n$-hexyl or $n$-octyl group.  See
the Supplementary Information.
\label{fig:Fig01}
}
\end{figure}

Organic electronics has emerged as competitive with silicon-based
electronics for many applications.  Reasons for this include ease
of manufacture and the possibility of printing of circuit components,
as well as the ability to use the principles of organic chemistry
to design the materials used in organic electronic devices at the
molecular level.  One particularly prominent example of organic
electronics is the increasingly wide spread use in displays of 
\marginpar{\color{blue} OSC}
organic light emitting diodes.  Another is the rapid emergeance of
organic solar cells (OSCs).  Once limited to niche markets because
of low photoconversion efficiencies (PCE), OSCs now have experimental
\marginpar{\color{blue} PCE}
PCEs as high as 18\% \cite{KGFN20,LJJ+20,LFI+20} with 20\% PCE
expected in the near future \cite{ZLQ+22}.  The typical modern OSC
is of the DON:ACC bulk heterojunction type where a donor material 
(DON) is mixed with an acceptor (ACC) material in an immiscible
\marginpar{\color{blue} DON,\\ ACC}
mixture.  Phase separation causes a dramatic increase in the surface
area of the DON:ACC heterojunction.  Such an OSC is a complex system,
the modeling of which would have to take into account a wide range
of time scales from femtoseconds to microseconds and length scales from
nanometers to centimeters and would have to include structural details
as well as exciton and charge transfer rates whose modeling is not
yet sufficiently well understood. Much less
ambitious, but still challenging, is to take an experimentally characterized
DON:ACC OSC and try to predict how chemical modifications will affect
the OSC PCE.  To this end, we have chosen to study a third generation
\marginpar{\color{blue} NFA}
nonfullerene acceptor (NFA) of the A-D-A form where A is an electron
accepting group and D is an electron donating group.  In particular,
IDIC-4Cl is well characterized with a PM6:IDIC-4Cl PCE of about 10\%.  
Changing 
\marginpar{\color{blue} A, D}
the Cl end groups is likely to have little effect other than on the
absorption spectrum and the highest occupied molecular orbital (HOMO)
\marginpar{\color{blue} HOMO,\\ LUMO,\\ FMO}
and lowest unoccupied orbital (LUMO) energies.  These frontier molecular
orbital (FMO) energies are measured by cyclic voltammetry (CV) and hence
\marginpar{\color{blue} CV} 
must be calculated as the appropriate oxidation and reduction potentials.
Finally these must be combined with other data to see how varying electron
donating and accepting groups in the A-D-A acceptor affect the DON:ACC
OSC PCE.  The molecules studied here are shown in {\bf Fig.~\ref{fig:Fig01}}.
Our approach is a first principles one using density-functional
\marginpar{\color{blue} DFT, TD} 
theory (DFT) and time-dependent (TD) DFT.  This allows us to obtain a
detailed picture of how molecular-level modifications
may change OSC performance at the macroscopic level.

The reader is referred to a recent review covering the rapid development 
of OSCs that has taken place over the last two decades \cite{ZLQ+22}.
That review divides the evolution of DON:ACC OSCs into three main stages
or generations.  During the first generation, typical OSCs were of the
P3HT:PC$_{61}$BM type where P3HT is a conducting polymer donor and PC$_{61}$BM
is buckminsterfullerene with an added substituent to improve solublity and
hence ease of manipulation.  The main emphasis during the first generation
was on optimizing the PCE through optimization of the morphology of the 
bulk heterojunction.  

The second generation consisted of replacing P3HT
with other DON polymers.  First principles modeling of OSC PCEs is not
discussed very much in the recent review \cite{ZLQ+22}.  Here we just
mention that the PCE is usually determined by using the Schockley diode
\marginpar{\color{blue} $J_{sc}$,\\ $V_{oc}$,\\ FF} 
model and measuring the short-circuit current density $J_{sc}$, open-circuit
voltage $V_{oc}$, and fill factor (FF), from which the PCE may be 
calculated.  Note however that the Schockley diode theory was developed
for inorganic devices and that the parameters entering into the theory
may need to be reinterpretted for organic devices \cite{GWWF10,GLW+10}, nor can
this theory describe the S-shaped current-voltage curves sometimes observed
for OSCs \cite{GWWF10,GLW+10,WGW+12,HPH14}.  Moreover modeling exciton and 
charge transfer requires deciding whether conduction is band-like or by 
hopping, when real organic systems often have elements of both 
\cite{GCE+19,GZC+20}.  In principle, electron and hole mobilities,
the absorption spectra of both the DON and
ACC phases, as the thickness of the layers, are needed
to correctly model the relation between photoabsorption and
photoconductivity spectra \cite{CD10,ISJ11,SSNC11,NKNN19}.  And there 
is the possibility that heterojunction surface states may also alter 
photoabsorption and photoconductivity spectra \cite{ISJ11}.  It was 
therefore fortunate that photon absorption and the associated exciton 
formation occured primarily in the ACC material in second generation
OSCs, allowing Scharber to 
propose a relatively simple partially phenomenological and partly detailed 
balance model for predicting OSC PCEs \cite{SMK+06,S16} [see the 
Supplementary Information (SI) for the present article].
\marginpar{\color{blue} SI} 

\begin{figure}
\begin{center}
\includegraphics[width=0.6\textwidth]{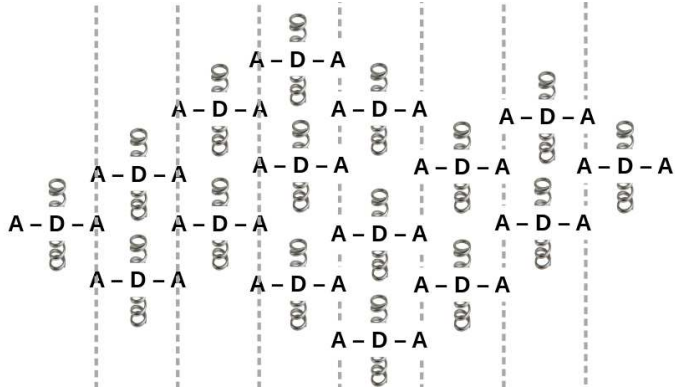}
\end{center}
\caption{
Schematic of the stacking of A-D-A molecules \cite{QCZ+18}.
Bulky out-of-plane side groups on the D units are represented by springs.
The $\pi$-stacking of A units is emphasized by joining stacked A units
with grey dotted lines.
\label{fig:stacking}
}
\end{figure}

The third generation is characterized by optimizing the ACC in the DON:ACC OSC.
Great success was found with small molecule NFAs 
\cite{LJZ+19,ZYC+20,YHG+20,LYW+21,ZLQ+22}, many with the A-D-A
configuration \cite{KMW+20,DKH+19,HZL+21,ZZH+21}.  
The greatest advantages of these NFAs over fullerenes are the
complementary donor/acceptor photoabsorption, improved solubility,
local crystallinity, and tunability of their optoelectronic 
properties \cite{KKS+21,YKI+21,LSL+21}.  These molecules
tend to be relatively flat with a large fused-ring D and strong
electron withdrawing units.  Their flatness allows them to stack,
forming locally crystalline layers as shown schematically in 
{\bf Fig.~\ref{fig:stacking}}.  Photon absorption and exciton formation
may occur in the D unit which then leads to charge separation with
D positively charged and A negatively charged.  Charge recombination
is minimized by stacking of A units in the crystal whose crystal 
structure allows the negative charges to conduct quickly away from 
the heterojunction.
Unfortunately too much local crystallinity can also be harmful,
so that bulky out-of-plane side groups (shown as springs in the
figure) are used to control the stacking density.  As we will show,
and provided we neglect any absorption features at the DON:ACC heterojunction
itself, photons
are being absorbed and excitons are now being formed in both the
DON and the ACC components, necessitating changes in the simple
Scharber model for predicting OSC PCEs.  In line with the philosophy
underlying Scharber's approach, we have presented a new finite-width
approach with only one additional assumption beyond Scharber's approach.  
Our description of
the A-D-A ACC would seem to indicate that the stronger the acceptor A, 
the better the OSC should work.  
We will argue that the opposite occurs in the present case.

The rest of the paper is organized as follows.  Computational details
are given in the next section.  Results are presented in Sec.~\ref{sec:results},
which is divided into three subsections.  Our choice of methodology is
carefully examined in the first subsection by comparing calculated results
with experimental results from the literature.  The next subsection focuses
on calculations characterizing the relative acceptor strength of the A units
and the consequences on physical properties.  The third and final subsection
predicts CV FMO values, absorption spectra, discusses the degree of charge
separation in the charge transfer excitations, and applies our finite-width
theory to predict OSC PCEs for our candidate molecules.
Section~\ref{sec:conclude} gives our concluding discussion.  
\section{Computational Details}
\label{sec:details}

The five molecules shown in Fig.~\ref{fig:Fig01} have been replaced by
somewhat simpler model compounds by replacing the four very flexible
$n$-hexyl (i.e., C$_6$H$_{13}$) chains in the molecules with methyl
groups.  The resultant molecules are much less flexible, making their
geometries easier to optimize.  The validity of this approximation will
be tested in Sec.~\ref{sec:results} and found to be acceptable.

Unless otherwise stated, all calculations were carried out using the 
{\sc Gaussian 09} program package \cite{g09}.  
Following the terminology of the Pople school, the triplet
functional/basis set/solvent defines our theoretical model chemistry (TMC).
\marginpar{\color{blue} TMC}
Geometries were optimized 
in chloroform using the integral equation formalism (IEF) variant of the implicit 
\marginpar{\color{blue} IEF, PCM}
solvent polarizable continuum model (PCM) (see Ref.~\cite{TMC05} for 
a review of PCM models) using DFT functional \cite{HS04} with the 6-311G(d,p) 
basis set \cite{KBSP80,MC80,FPH+82}.  After testing three hybrid functions 
(B3LYP \cite{SDCF94}, PBE0 \cite{AB99}, B3PW91 \cite{PW92}) and a range-separated 
hybrid functional (HSEH1PBE) \cite{HS04} for the geometry optimization of {\bf M$_0$}, 
it was decided to use HSEH1PBE for all of the molecules.

Start geometries for the optimizations were constructed using the {\sc Gauss View} 
program \cite{GaussView}.  {\sc Molden} \cite{molden} was also used for visualizing
molecules.  Vibrational frequencies were calculated to verify that 
the optimized geometries are true minima.  Optimized geometries and the lowest
vibrational energies are given in the SI.  Translational
and rotational energies often come out as imaginary but of small magnitude 
(less than 11.0 cm$^{-1}$) because of numerical approximations.  

Once their geometries were optimized, these same model molecules were 
used to calculate all of the properties reported here.  Note that we will
show that the replacement of the four $n$-hexyl alkyl groups by simpler methyl 
groups has little  direct effect on the electronic properties of these molecules. 
\marginpar{\color{blue} X}
However the large flexible side groups (X) do increase solubility and help to
control local aggregation in the solid state.  Such local aggregation is important
because it may help to facilitate the transformation of a charge transfer state at 
the DON:ACC interface into fully separated charges, but may also also have a 
negative effect on OSC flexibility and morphology.  In particular, a very similar molecule 
with $p$-hexylphenyl groups rather than $n$-hexyl groups is known to have a crystal structure
typical of ideal J-aggregation \cite{QCZ+18}, meaning that absorption spectra
are expected to be red shifted and to show additional peaks due to intermolecular 
stacking interactions (see, e.g., Ref.~\cite{DCJ+18} for a brief history and review 
of excitonic effects on spectra).

CV redox potentials were also calculated in acetonitrile as
this is closer to the preferred solvent used in the experimental measurements.
Outside of the field of materials for electronics applications, CV data is simply 
described as measuring oxidation and reduction potentials.  However, within the 
field of materials for electronics applications, these redox
potentials are usually reported as frontier molecular orbital (FMO) energies. 
[An exception is a University-level general chemistry textbook that also
discusses materials for electronics applications (see page 779 of 
Ref.~\cite{OGC08}).]
The identification of CV redox potentials as FMO energies is 
misleading \cite{PG18},
causing some researchers to misidentify DFT HOMO and LUMO
energies with CV HOMO and CV LUMO energies. 
Proper calculation of CV FMO energies involves calculating the Gibbs' free energy for ionization and electron attachment 
in solution \cite{NLC10}.  However it is a reasonable approximation to ignore entropic and pressure
volume effects and just calculate ionization potentials (IPs) and electron
\marginpar{\color{blue} IP, EA}
affinities (EAs) in solution.  As we shall see, we do obtain CV FMO energies with 
this method which agree reasonably well with experiment.

Sometimes though, we would like to compare experimental CV FMO energies directly
with DFT FMO energies, but is this really reasonable?  The answer is that we
{\em can} use the DFT FMO energies as approximate CV FMO energies
{\em provided} we are very careful about how we do this.  The reason goes
back to the basics of Kohn-Sham DFT \cite{KS65} itself.  For reasons
of simplicity and rigor we will specialize to the HOMO energy.
In the original theory, the exchange-correlation potential $v_{xc}$ was
a purely multiplicative function for which the HOMO energy would be
minus the ionization potential ($E_{\mbox{HOMO}}=-\mbox{IP}$),
provided $v_{xc}$ is exact.  (see, e.g., Ref.~\cite{C99} and references
therein).  However it was later discovered that the exact $v_{xc}$ contains
a particle-number derivative discontinuity \cite{PL83,SS83} which
\marginpar{\color{blue} LDA, GGA}
is absent in both the local density approximation (LDA) and in generalized
gradient approximation (GGA) functionals, leading to the HOMO energy
to be closer to minus the average of the ionization potential and the
electron affinity ($E_{\mbox{HOMO}}=-(\mbox{IP}+\mbox{EA})/2$) within
a reasonable parabolic approximation.  This behavior is in marked contrast
to the familiar Koopmans' interpretation of Hartree-Fock molecular orbital
energies that $E_{\mbox{HOMO}} \approx -\mbox{IP}$ and $E_{\mbox{LUMO}}
\approx -\mbox{EA}$ \cite{K34}.  It also means that the introduction of
global and range-separated functionals with part exact exchange is
no longer Kohn-Sham DFT, even if the total energies are (to within
a localization of $v_{xc}$) DFT energies.  In particular, varying
the functional, and particularly how the exact exchange enters the functional,
provides a way to {\em tune} the orbital energies.  In fact, the concept of
optimal tuning was introduced in TD range-separated hybrid
DFT as a way to fix the range-separation parameter while simultaneously
improving charge-transfer excitation energies \cite{BLS10}.  In so far
as DFT orbital energies are much more sensitive to the choice of functional
than are DFT total energies, tuning the functional can allow the DFT
FMO energies to approach CV FMO energies calculated using the DFT
$\Delta$SCF procedure with the same functional.  In a somewhat similar
spirit, and knowing that DFT orbital energies and $\Delta$SCF energies
have been frequently noted to be linearly correlated
(see, e.g., Refs.~\cite{HCS02,HDCS02,DCT+15}), we have tried several 
different functionals to see which, if any of them, give DFT FMO energies 
closest to experimentally obtained CV FMO energies.  (See Sec.~\ref{sec:results}.)

TD-DFT absorption spectra were calculated in chloroform using the same\\
HSEH1PBE/6-311G(d,p) level of calculation. 
The absorption spectra have been calculated for 30 
singlet-singlet transitions and were gaussian-convoluted before comparing
with experimental spectra.  The {\sc SWIZARD} program \cite{G13}
was used to help assign the calculated transitions.

DON:ACC PCEs were estimated using a simple home-made {\sc Python} program for
making Scharber plots designed to vary DON at constant ACC (see the SI) which 
was modified as further described in Sec.~\ref{sec:results} to be able to vary
ACC at constant DON.

Charge transfer within our A-D-A ACC molecules was analyzed using transition
density matrix (TDM) heatmaps made using version 3.1 of the {\sc Multiwfn} 
program \cite{LC12,Multiwfn}.
\marginpar{\color{blue} TDM}
This is also further described in Sec.~\ref{sec:results}.

\section{Results}
\label{sec:results}

Our results are divided into three parts:  (1) validation of our choice
of TMC---namely, the HSEH1PBE/6-311G(d,p) with the
PCM implicit solvent model for chloroform---by comparison against experimental
values for IDIC-4Cl;  (2) use of various analytical tools to characterize 
chemical notions such as relative donor/acceptor abilities and chemical 
reactivity for our four candidate molecules; and finally (3) concrete
physical predictions regarding the physical properties of the candidate
molecules.

\subsection{Theoretical Model Chemistry Validation}
\label{sec:TMCA}

We first validate our TMC against reported experimental
results for IDIC-4Cl.  A caveat is that there are several different IDIC-4Cl
molecules reported in the literature (see the SI), 
albeit with similar electronic properties.  Our simplified model [IDIC(tetramethyl)-4Cl] is 
expected to be able to reproduce the properties of all of these different
IDIC-4Cl molecules, to the extent that results are independent of the side
chains (X).

\begin{figure}
\begin{center}
\includegraphics[width=0.8\textwidth]{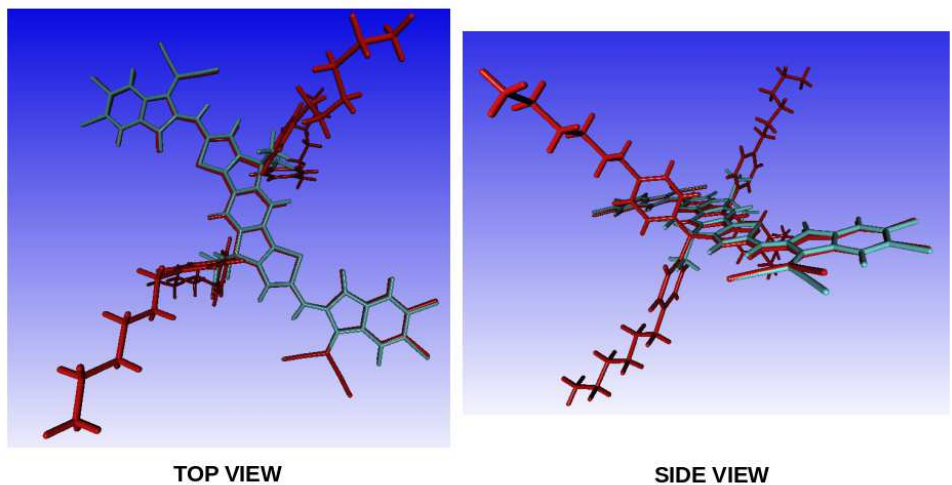}
\end{center}
\caption{
Comparison of the single-crystal X-ray geometry of 
IDIC(tris(4-hexylphenyl))-4Cl (Cambridge Crystallographic Data Centre
database identifier YISJOC, deposition number 1860549, Ref.~\cite{QCZ+18}) 
in red with our HSEH1PBE/6-311G(d,p)/PCM(chloroform) optimized geometry for 
IDIC(tetramethyl)-4Cl in green.
\label{fig:superposition}
}
\end{figure}
We were able to find a single-crystal X-ray geometry for IDIC-4Cl, but only
for IDIC(tris(4-hexylphenyl))-4Cl \cite{QCZ+18}.  
{\bf Figure~\ref{fig:superposition}} shows a superposition of this structure
with our HSEH1PBE/6-311G(d,p)/PCM(chloroform) optimized geometry for 
IDIC(tetramethyl)-4Cl using {\sc Molden} \cite{molden}.  Except for the side
chains, the structures are remarkably well matched.  The $(x,y,z)$-coordinates
of  HSEH1PBE/6-311G(d,p)/PCM(chloroform) optimized IDIC(tetramethyl)-4Cl
geometry are given in the SI.

CV data are routinely reported for molecules which are 
important for organic electronics.  
CV FMO energies are obtained
by measuring the redox potentials for the molecule in question and for
ferrocene under the same conditions, correcting the redox potential of the
molecule using ferrocene redox potential, and then adding an additional 
correction to put the CV HOMO and CV LUMO on the absolute vacuum scale.


\begin{table}

\begin{center}
\begin{tabular}{ccccc}
\hline \hline
X & CV HOMO Energy & CV LUMO Energy & CV H-L Gap & Reference \\
\hline 
\multicolumn{5}{c}{IDIC(X)}\\
tetrakis(4-hexylphenyl) & -5.70 eV & -3.87 eV & 1.83 eV & \cite{QCZ+18} \\
tetraoctyl              & -5.77 eV & -3.98 eV & 1.78 eV & \cite{HZLZ+21} \\
tetrahexyl              & -5.69 eV & -3.91 eV & 1.78 eV & \cite{LHZ+16} \\
\multicolumn{5}{c}{IDIC(X)-4Cl}\\
tetrakis(4-hexylphenyl) & -5.78 eV & -4.02 eV & 1.76 eV & \cite{QCZ+18} \\
tetraoctyl              & -5.88 eV & -3.98 eV & 1.90 eV & \cite{WDZ+20,ZWD+20} \\
tetraoctyl              & -5.85 eV & -4.05 eV & 1.80 eV & \cite{HZLZ+21} \\
tetrahexyl              & -5.81 eV & -4.01 eV & 1.80 eV & \cite{LMS+19} \\
tetramethyl             & -5.99 eV & -3.89 eV & 2.10 eV & $\Delta$SCF \\
tetramethyl             & -4.81 eV & -4.04 eV & 1.77 eV & ``Koopmans'' \\
\hline \hline
\end{tabular}
\end{center}

\caption{Comparison of experimental and calculated (PW) CV FMO values.
The experimental CV FMO energies are typically measured in acetylnitrile  
with ($n$-Bu)$_4$NPF$_6$ (0.1 M) as supporting electrolyte, Pt wire as 
counter electrode, and Ag/Ag$^+$ as reference electrode.  Our calculations
are for the HSEH1PBE/6-311G(d,p)/PCM(chloroform) theoretical model chemistry.
\label{tab:FMOCl}}

\end{table}


{\bf Table~\ref{tab:FMOCl}} shows measured CV FMO energies for several 
IDIC and IDIC-4Cl molecules taken from the literature.  IDIC(tetrahexyl)
and IDIC(tetraoctyl) have nearly identical CV FMO energies.  Similarly
IDIC(tetrahexyl)-4Cl and IDIC(tetraoctyl)-4Cl have nearly identical
CV FMO energies, consistent with the idea that the hexyl and octyl groups
are chemically similar, neither donating nor accepting electrons.  This
comforts us in our expectation that IDIC(tetramethyl)-4Cl should also have
very similar CV FMO energies. 

Interestingly, as the CV HOMO and CV LUMO energies of 
IDIC(tetrakis(4-hexylphenyl)) are greater than those of 
IDIC(tetraoctyl), the 4-hexylphenyl should also be a (+M) electron donor.  
However, the experimental CV HOMO and CV LUMO energies of\\
IDIC(tetrahexyl) are very similar to those of 
IDIC(tetrakis(4-hexylphenyl)).  This may simply be a question of
differences in how the CV FMO energies were measured 
and how accurately they could be determined experimentally
since the  CV HOMO energy of
IDIC(tetrakis(4-hexylphenyl))-4Cl is higher than that of 
IDIC(tetrahexyl)-4Cl and of IDIC(tetraoctyl)-4Cl.  Thus the
classification of 4-hexylphenyl as a +M donor seems reasonable.
Note that the experimental CV LUMO energies are all very similar 
for the three IDIC-4Cl molecules.  


\begin{table}

\begin{center}
\begin{tabular}{cccccc}
\hline \hline
                 & B3LYP$^a$ & PBE0$^b$  & B3PW91$^c$ & HSEH1PBE$^d$ & Expt.$^e$ \\
\hline 
LUMO             & -3.79 & -4.25 & -3.87  & -4.04    & -4.01 \\
HOMO             & -5.93 & -5.47 & -6.04  & -5.81    & -5.81 \\
$E_{\mbox{H-L}}$ &  2.14 &  1.22 &  2.17  &  1.77    &  1.79 \\
\hline \hline
\end{tabular}
\end{center}
\caption{The energy levels of HOMO (eV), LUMO (eV), and the energy gap 
$E_{\mbox{H-L}}$ (eV) for the acceptors molecule obtained by DFT/6-311G(d,p) 
in chloroform solution.
$^a$ Ref.~\cite{SDCF94}.
$^b$ Ref.~\cite{AB99}.
$^c$ Ref.~\cite{B93}.
$^d$ Ref.~\cite{HS04}.
$^e$ CV FMO energies for IDIC(tetrahexyl)-4Cl from Ref.~\cite{LMS+19}.
\label{tab:tab1}}

\end{table}

\begin{figure}
\begin{center}
\includegraphics[width=0.8\textwidth]{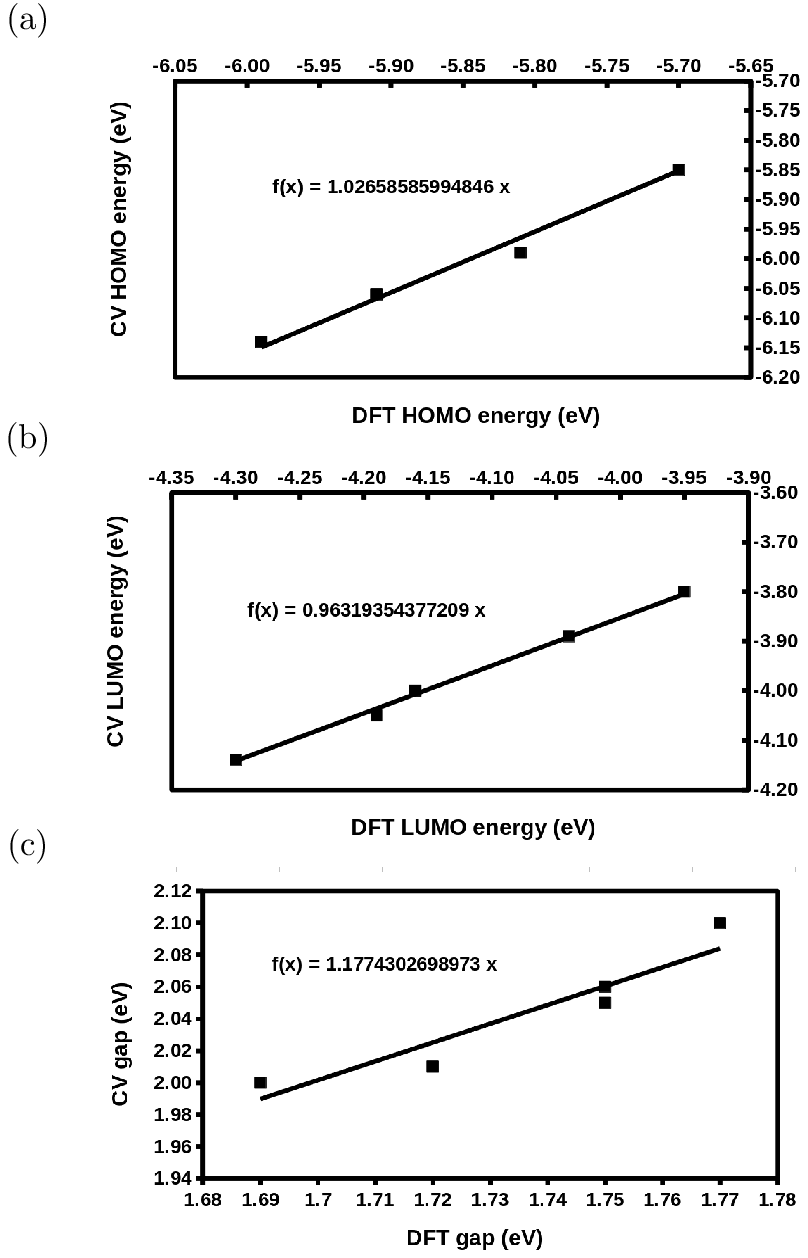}
\end{center}
\caption{Forced intercept least squares fit between DFT FMO energies and 
CV ($\Delta$SCF) FMO energies calculated using the 
HSEH1PBE/6-311G(d,p)/PCM(chloroform) theoretical model chemistry.  
\label{fig:OptimalTuning}
}
\end{figure}
Calculated $\Delta$SCF CV FMO energies are shown in Table~\ref{tab:FMOCl}.
In Sec.~\ref{sec:details}, we raised the question of which functionals give
DFT FMO energies closest to the experimental CV FMO energies?
{\bf Table~\ref{tab:tab1}} shows that\footnote{Note that B3LYP here is the original {\sc Gaussian} B3LYP
functional, rather than the ``corrected'' B3LYP functional which is the 
default in the {\sc Turbomol} and {\sc Orca} programs and which gives very
different total energies.  The latter B3LYP functionals
from the original {\sc Gaussian} version by replacing {\sc Gaussian}'s 
VWN RPA LDA parameterization with the original VWN Monte Carlo LDA 
paramerization of Vilko, Wilk, and Nusair \cite{VWN80} (called VWN5) in 
{\sc Gaussian}.}, of the functionals tested, the 
HSEH1PBE functional meets this criterion best. 
In the end, because of the fit to experiment, we 
think that our DFT FMO energies might even be a better predictor of 
experimental CV FMO energies, although we prefer to report CV FMO energies
calculated using the $\Delta$SCF procedure.

\begin{figure}
\begin{center}
\includegraphics[width=0.8\textwidth]{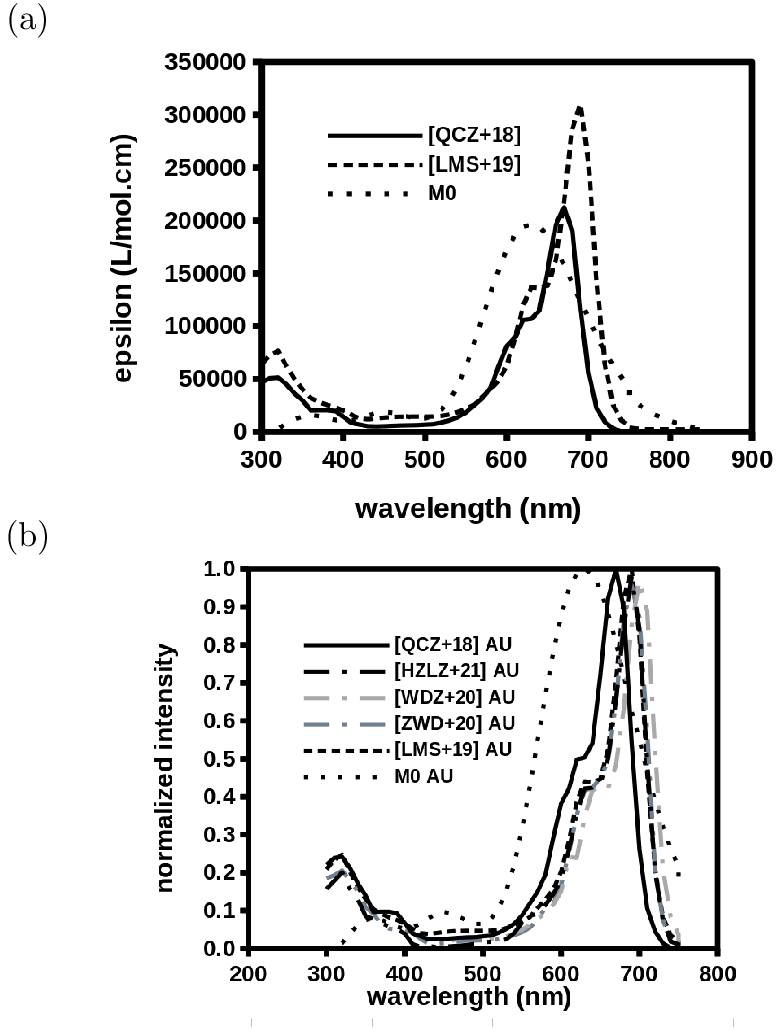}
\end{center}
\caption{
Absolute (a) and renormalized (b) IDIC(X)-4Cl absorption spectra from 
the literature and calculated using TD-DFT within the 
HSEH1PBE/6-311G(d,p)/PCM(chloroform) theoretical model chemistry:  
[QCZ+18], X = tetrakis(4-hexylphenyl), 
in chloroform \cite{QCZ+18};
[WDZ+20], [ZWD+20], [HZLZ+21], X = tetraoctyl, 
in chloroform \cite{WDZ+20,ZWD+20,HZLZ+21};
[LMS+19], X = tetrahexyl, in chloroform \cite{LMS+19}; and
{\bf M$_0$}, X = tetramethyl, present calculations.
\label{fig:IDIC-4Clspectra}
}
\end{figure}
Absorption spectra are another type of data which are routinely reported 
for molecules which are important for organic electronics.  Care must be
taken in measuring solution spectra because flat polyaromatic molecules,
such as substituted IDICs are likely to form red-shifted J-aggregates 
\cite{QCZ+18}.  We will come back to this below.  In general
IDIC-Cl spectra are found to be red shifted with respect to the corresponding
IDIC spectra \cite{QCZ+18,HZLZ+21}, consistent with the previous observation that
chloro groups tend to close the CV HOMO-LUMO gap. 
{\bf Figure~\ref{fig:IDIC-4Clspectra}} shows literature spectra for 
IDIC(X)-4Cl molecules digitized using the interactive internet-based
{\sc WebPlotDigitizer} \cite{wpd} and combined into a single diagram.
It is immediately clear from the figure that there is no significant difference 
between IDIC(tetrahexyl)-4Cl and IDIC(tetraoctyl)-4Cl which suggests that
we are justified in using our IDIC(tetramethyl)-4Cl model.  However
the IDIC(tetrakis(4-hexylphenyl))-4Cl spectrum is blue shifted 
with respect to the other IDIC-4Cl spectra.  This suggests an opening
of the IDIC-4Cl CV HOMO-LUMO gap in the case of IDIC(tetrakis(4-hexylphenyl))-4Cl.
It may also be that some aggregate formation was occuring in the solutions
for which the other IDIC-4Cl spectra were measured.  

Our calculated TD-DFT spectrum is also shown in Fig.~\ref{fig:IDIC-4Clspectra}.
Following usual practice, the calculated spectrum has been gaussian-convoluted.
In this case, the full-width at half maximum (FWHM) of the gaussian that was
\marginpar{\color{blue} FWHM}
used is a bit too broad compared with the measured peak widths.  This is probably
also why no shoulder is observed in the theoretical spectrum while a short
wavelength shoulder is quite evident on the main peak in the experimental
spectrum.  More importantly the maximum of the main peak 
in the TD-DFT calculation is red shifted compared to the experimental
IDIC(tetrahexyl)-4Cl peak 
by about 0.2 eV, which is within the expected limit of the computational
method.

A third type of data reported in this field is the very {\em raison d'\^etre}
for the field.  These are the performance parameters of actual
organic photovoltaic cells determined from current density $J$ versus
voltage $V$ curves.  Such curves have the general form 
predicted by some variant of the Shockley diode equation \cite{S50,S01},
\begin{equation}
  J = J_0 \left[ 1-\exp \left( \frac{V+IR}{n V_T} \right) \right] +
  J_{ph} \, ,
  \label{eq:Shockley}
\end{equation}
where $V_T = k_B T/e$ and $J_{ph}$ is the photocurrent.  Rather than
try to invert the equation to obtain $J_0$, $R$, and the ideality factor $n$,
it is traditional to report another set of parameters to characterize
photovoltaic cell performance.  These are the 
$V_{oc}$,
$J_{sc}$, and FF. 
$V_{oc}$ is related to the difference between the
difference between the CV LUMO energy of the acceptor and 
the CV HOMO energy of the donor which provides the driving force
for charge separation at the donor-acceptor interface.  Empirically,
for some solar cells, with ohmic contacts, it has been found that,
\begin{equation}
   eV_{oc} = E_{\mbox{CV LUMO}}^{ACC} - E_{\mbox{CV HOMO}}^{DON} -\delta \, .
   \label{eq:Voc}
\end{equation}
The $\delta$ correction likely arises for several physical reasons. 
It is already found in the Shockley diode equation itself \cite{SMK+06,S16}.  
However additional contributions may arise due to band bending in the electrodes 
\cite{MBHR03} or from the simple fact that the values of the CV FMO energies used
in the equation are obtained under different conditions than those in
actual solar cells.  $J_{sc}$ 
is related to
how much light is converted into electrical energy, and so it is ultimately
also related to the absorption spectrum.  The FF 
measures how close the J-V curve is to a perfect rectangle.  
Physically the FF is
a monotonically increasing function of $V_{oc}/nV_T$ \cite{R22}.  High
values of FF are favored by large open-circuit voltages and low values
of the ideality factor $n$.  The value of the ideality factor is closely
related to morphology as low values of $n$ mean that charges travel
farther before electron-hole recombination occurs. Together, these
allow the calculation of the PCE  $\eta$ via 
the equation,
\begin{equation}
  \eta = \frac{J_{sc} V_{oc} \mbox{FF}}{P_s} \, ,
  \label{eq:PCE}
\end{equation}
where $P_s$ is the solar power.  Under AM 1.5G conditions, 
\marginpar{\color{blue} $P_s$}
\begin{equation}
  P_s = \int_0^\infty F(\lambda) \, d\lambda = \mbox{99.971 mW/cm$^2$} \, ,
  \label{eq:power}
\end{equation}
where $F(\lambda)$ is the irradiance (in W/m$^2$.nm).


\begin{table}

\begin{center}
\begin{tabular}{cccccc}
\hline \hline
CV HOMO Energy & CV LUMO Energy & $V_{oc}$ & $J_{sc}$ & FF &
PCE \\
\hline 
\multicolumn{6}{c}{PBDB-T2Cl:IDIC(tetrakis(4-hexylphenyl))-4Cl \cite{QCZ+18}}\\
-5.51 eV & -3.57 eV & 0.83 V & 16.21 mA/cm$^2$ &
                        68.69\% & 9.24\%  \\
\multicolumn{6}{c}{BSCl:IDIC(tetraoctyl)-4Cl \cite{ZWD+20}}\\
-5.55 eV & -3.30 eV & 0.865 V & 21.5 mA/cm$^2$ &
                        70.0\% & 13.03\% \\ 
\multicolumn{6}{c}{BSCl-C1:IDIC(tetraoctyl)-4Cl \cite{WDZ+20}}\\
-5.59 eV & -3.34 eV & 0.564 V & 4.9 mA/cm$^2$ &
                        33.9\% & 0.90\% \\ 
\multicolumn{6}{c}{BSCl-C2:IDIC(tetraoctyl)-4Cl \cite{WDZ+20}}\\
-5.58 eV & -3.39 eV & 0.865 V & 20.1 mA/cm$^2$ &
                        71.3\% & 12.40\% \\ 
\multicolumn{6}{c}{BSCl-C3:IDIC(tetraoctyl)-4Cl \cite{WDZ+20}}\\
-5.55 eV & -3.37 eV & 0.870 V & 14.2 mA/cm$^2$ &
                        67.0\% & 8.25\% \\ 
\multicolumn{6}{c}{ZR1-Cl:IDIC(tetraoctyl)-4Cl \cite{HZLZ+21}}\\
-5.51 eV & -3.60 eV & 0.87 V & 18.30 mA/cm$^2$ &
                        68.03\% & 10.81\% \\ 
\multicolumn{6}{c}{PM6:IDIC(tetrahexyl)-4Cl \cite{LMS+19}}\\
-5.50 eV & -3.61 eV & 0.767 V & 17.87 mA/cm$^2$ & 74.76\% & 10.25\% \\
\hline \hline
\end{tabular}
\end{center}

\caption{Photovoltaic properties for some donor:IDIC(X)-4Cl solar cells.
The CV FMO energies are for the donor.  The reader is referred to the
orginal articles for an explanation of the donor molecule name.
\label{tab:PVCl}}

\end{table}

\begin{figure}
\begin{center}
\includegraphics[width=0.6\textwidth]{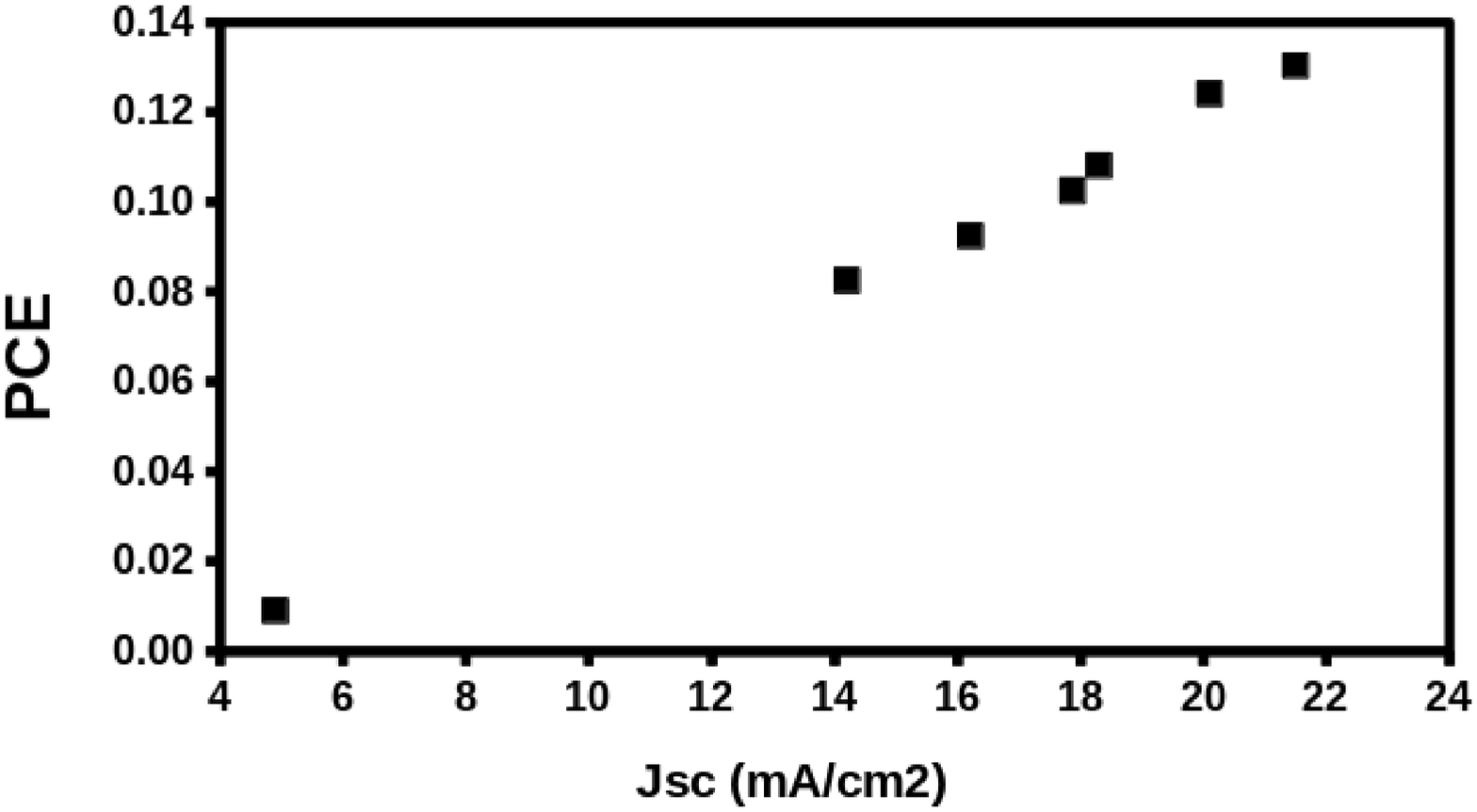}
\end{center}
\caption{
Graph of the data from Table~\ref{tab:PVCl} showing the near proportionality
of the photoconversion efficiency and the open-circuit current.
\label{fig:PCEvsJsc}
}
\end{figure}
{\bf Table~\ref{tab:PVCl}} shows actual experimental data taken from
the literature.  {\bf Figure~\ref{fig:PCEvsJsc}} shows that, for this
data, the PCE is nearly proportional to $J_{sc}$.  This is because
both the FF and $V_{oc}$ are essentially the same for all of the 
photocells.  Except for the somewhat aberrant BSCl-Cl:IDIC(tetraoctyl)-4Cl
photocell, the value of the FF varies from 67.0\% to 74.76\% with an 
average value of 69.96\%.  The values of $V_{oc}$ are too similar to 
confirm an equation of the form of Eq.~(\ref{eq:Voc}).  Instead, we assume 
the form of the equation and calculate that the average value of 
$\delta$ = 0.74 eV.  

\begin{figure}
\begin{center}
\includegraphics[width=0.6\textwidth]{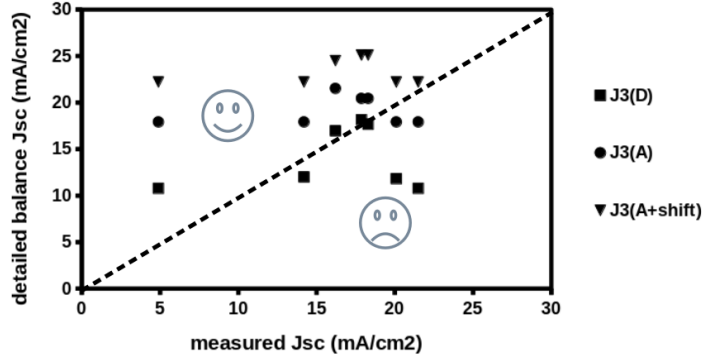}
\end{center}
\caption{
Detailed balance short-circuit currents calculated from the
acceptor CV gap with and without red shift and from the donor
CV gap.  In this diagram A stands for ACC and D stands for DON.
\label{fig:Jdb}
}
\end{figure}
But how can we estimate the short-circuit current?  The external 
quantum efficiency $\mbox{EQE}(\lambda)$ is the percent of photons
\marginpar{\color{blue} EQE}
converted into current.  It can be shown that,
\begin{equation}
  J_{sc} = \int_0^\infty \mbox{EQE}(\lambda) F(\lambda) \lambda \, d\lambda
  \, .
  \label{eq:EQE}
\end{equation}
Shockley and Queisser introduced the detailed balance (db) model which assumes 
\marginpar{\color{blue} db}
that every photon whose energy exceeds the CV gap is absorbed and forms 
an electron-hole pair \cite{SQ61}.  Then
\begin{equation}  
   \mbox{EQE}(\lambda) =  \theta(\lambda_g - \lambda) \, ,
   \label{eq:db.1}
\end{equation}
where $\theta$ is the Heaviside (step) function, so that
\begin{equation}
  J_{db}(E_g) = \int_0^{\lambda_g} F(\lambda) \lambda \, d\lambda \, ,
  \label{eq:db.2}
\end{equation}
where $\lambda_g = hc/E_g$ is the wavelength corresponding to the CV
gap.  An adequate approximation \cite{ARE+15} to the experimental 
AM 1.5G irradiance \cite{NREL} is given in Ref.~\cite{ARE+15} and 
\marginpar{\color{blue} AM 1.5G}
leads to,
\begin{equation}
  J_{db}(E_g) = \left( \mbox{73.531 mA/cm$^2$} \right)
  e^{-(0.440)(E_g/\mbox{eV})^{1.8617}} \, .
  \label{eq:Jdb}
\end{equation}
This allows us to obtain an upper limit on the PCE of the solar
cells whose parameters are given in Table~\ref{tab:PVCl}.  Unlike
organic solar cells with fullerene acceptors, it is no longer a 
good assumption that absorption and exciton formation takes place
only in the DON.  We must also test Eq.~(\ref{eq:Jdb}) for
the ACC.  (Note that we are ignoring absorption by heterojunction
surface states, which seems a reasonable first approximation.)
The detailed balance $J_{sc}$ 
\marginpar{\color{blue} ACC, DON}
calculated for the acceptor and for the donor are compared with 
the measured $J_{sc}$ 
in {\bf Fig.~\ref{fig:Jdb}}.
Anything below the dashed line (frowny face region) represents
a violation of energy conservation.  This occurs both for the
donor and for the acceptor, though it is worse for the donor.
However it must be kept in mind that there is a significant
red shift in the absorption spectrum in going from solution
to the solid state (see SI-Spectral Shift).  Subtracting this
shift from the ACC CV gap restores energy conservation (smiley
face region), indicating the importance of this small shift.
The shifted acceptor point which is closest to the dashed line
is the BSCl:IDIC(tetraoctyl)-4Cl solar cell whose 13.03\%
PCE is essentially the same as the detailed balance limit for
this type of solar cell.

\begin{figure}
\begin{center}
\includegraphics[width=0.6\textwidth]{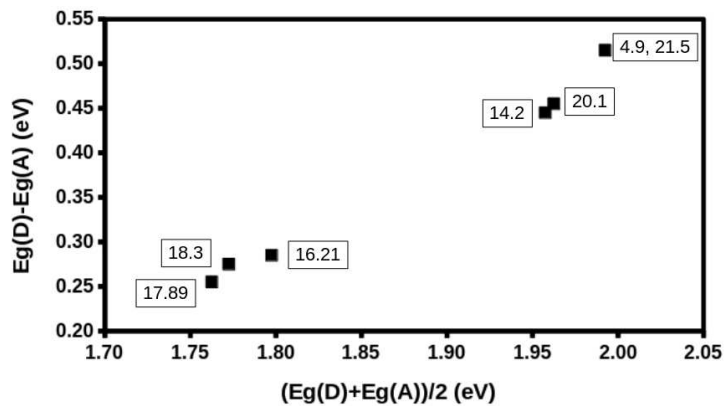}
\end{center}
\caption{
Gap difference versus average gap. The framed numbers are
measured short-circuit currents in mA/cm$^2$.  Note that the
acceptor CV gaps include the red shift correction.
\label{fig:EaveVSdeltaE}
}
\end{figure}
In previous work, Scharber constructed a contour plot of 
DON:PCBM bulk heterojunction solar cell PCEs as a function of
the DON CV LUMO energy and the donor CV gap \cite{SMK+06,S16}.  
It is an attempt to make an educated guess as to the best obtainable
PCE for different DON polymers.  This is, of course, tricky 
because it refers not to every solar cell, but only to the ones with
optimal ohmic contacts, good morphology, etc.  Scharber's theory 
is very similar to what we have been doing here, except that 
Scharber assumed that all excitons are formed in the DON, which
is reasonable for DON:PCBM solar cells.  Here we see that excitons 
must also be formed in the ACC, hence necessitating a change 
in Scharber's theory.  Like Scharber, we will search for a description
of the best $J_{sc}$ 
consistent with the known
DON and ACC band gaps.  {\bf Figure~\ref{fig:EaveVSdeltaE}}
shows that the difference between the DON and ACC CV gaps
is nicely correlated with the average of the DON and ACC 
CV gaps.  That is because the ACC CV gap is relatively 
constant compared to the DON CV gap.  More importantly, we
see that there is a general tendency for $J_{sc}$ to increase
as the band gap difference increases, {\em provided} we ignore the
lower-performing solar cells with $J_{sc}$ equal to 4.9 mA/cm$^2$
and to 14.2 mA/cm$^2$.  This is justifiable because accurate benchmarking
of organic solar cell performance requires a rigor which is often lacking
in exploratory results reported in the literature.  Certified performance
numbers require very carefully prepared standard cells and must be 
performed by an independent third-party laboratory. Moreover, 
to quote Ref.~\cite{ZLQ+22}, ``It is important
to note that not all independent measurements performed by third-party
laboratories are accepted for entry into the Solar Cell Efficiency
Tables, as only nine laboratories are officially recognized.''
Most of the time, this means that the performance data
reported in the literature is not quantitatively reproducible and is often
on the low side.  Hence we feel justified in ignoring the solar cell with
$J_{sc}$ equal to 4.9 mA/cm$^2$.  Ignoring the solar cell with $J_{sc}$ equal 
to 14.2 mA/cm$^2$ is more debatable, but we feel that this experimental value 
is probably also underestimated.  Of course, we could also have ignored
the solar cells with $J_{sc}$ equal to 16.21 mA/cm$^2$, 17.89 mA/cm$^2$,
and 16.21 mA/cm$^2$, but we have chosen not to do so.  Our theory 
here is that better performance is obtained by having absorption
occurring at different wavelengths in the ACC and in the DON.

\begin{figure}
\begin{center}
\includegraphics[width=0.6\textwidth]{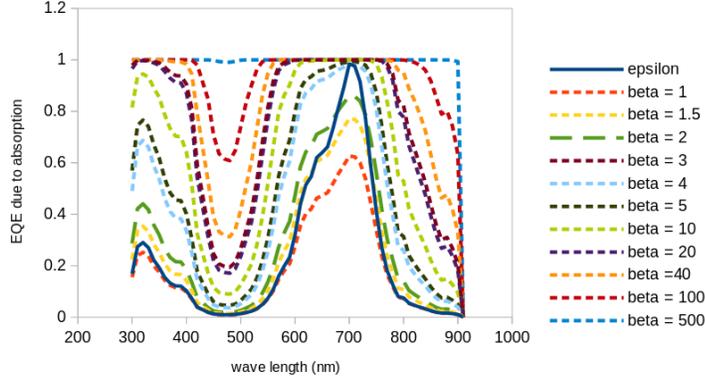}
\end{center}
\caption{
Graph of the function $\mbox{EQE}(\lambda) = 
1 - \exp \left( -\beta \bar{\epsilon}(\lambda) \right)$ for different values
of $\beta$ for IDIC-4Cl(tetrakis(4-hexylphenyl)).  
\label{fig:EQE}
}
\end{figure}
Let us be more precise.  If absorption is included in the EQE,
then the function becomes \cite{ARE+15}
\begin{equation}
  \mbox{EQE}(\lambda)  =  1 - e^{-\alpha(\lambda) \ell} 
   = 1 - e^{-\beta \bar{\epsilon}(\lambda)} \, .
  \label{eq:EQEtoo}
\end{equation}
Here $\alpha(\lambda)$ is the attenuation coefficient in the Bouguer-Lambert 
law and $\ell$ is the path length.  For a dilute solution, we recover
Beer's law,
\begin{equation}
   \alpha(\lambda) = \ln(10) \epsilon(\lambda) c  \, .
   \label{eq:Beer}
\end{equation}
As organic solar cells are {\em not} dilute, Beer's law is only indicative
and this is why the absorption coefficient for organic solids is reported
in terms of a normalized quantity,
\begin{equation}
  \bar{\epsilon}(\lambda) = \frac{\epsilon(\lambda)}{\epsilon_{\mbox{max}}}
  = \frac{\epsilon(\lambda)}{\epsilon_{\mbox{max}}} \, .
  \label{eq:normalized}
\end{equation}
Evidently for a dilute solution,
\begin{equation}
  \beta = \ln(10) \epsilon_{\mbox{max}} c \ell \, .
  \label{eq:beta}
\end{equation}
{\bf Figure~\ref{fig:EQE}} compares $\bar{\epsilon}(\lambda)$ with this
$\mbox{EQE}(\lambda)$ for different values of $\beta$ for the particular
case of IDIC-4Cl(tetrakis(4-hexylphenol)).  The CV gap is 1.76 eV which
corresponds to a gap wave length of 705 nm which explains the position
of the main peak in the absorption spectrum.  The function 
$\mbox{EQE}(\lambda)$ is seen to be qualitatively similar to the absorption
spectrum for $\beta \leq 2$.  This is fine for very thin films.  However
as the path length increases for normal thicknesses, $\beta$ increases
and becomes more and more like a square function (case of $\beta = 500$).
However, even for $\beta = 100$, we see a dip in $\mbox{EQE}(\lambda)$
near 450 nm. So much for the ACC.

Now consider the DON.  It is expected to have
an absorption peak at around the wave length corresponding to its CV gap.
In the case of the highest performing solar cell, the donor is BSCl
and its gap wave length is 551 nm which is expected to lead to an
$\mbox{EQE}(\lambda)$ which will nicely fill in the dip in the ACC 
$\mbox{EQE}(\lambda)$ function.  In contrast, the gap wave length of the
PM6 solar cell is 656 nm whose $\mbox{EQE}(\lambda)$ is less likely to
be able to fill in the dip in the ACC $\mbox{EQE}(\lambda)$ function.
This explains the perhaps counter-intuitive result that increasing the 
donor gap may actually lead to an increased short-circuit current and 
a larger PCE rather than to a decreased short-circuit current and a
smaller PCE.

\begin{figure}
\begin{center}
\includegraphics[width=0.8\textwidth]{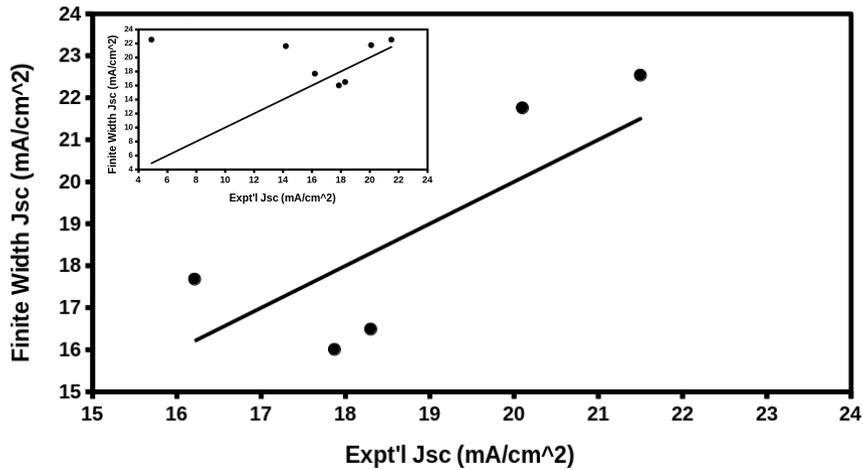}
\end{center}
\caption{
Comparison of fw model results with experimentally measured open-circuit
currents using $\Delta \lambda = \mbox{200 nm}$ and DON CV FMO energies
from Table~\ref{tab:PVCl} as well as a DON gap of of 1.78 eV for IDIC(tetrahexyl) and 
1.80 eV for the other IDIC based upon data in Table~\ref{tab:CDFT}.
The inset shows the two data points that we have chosen to ignore on the
grounds that they are likely underestimates of cell performance (see text).
In both the main graph and the inset, the line represents perfect agreement
between the model predictions and experimental measurements.
\label{fig:FiniteWidthJsc}
}
\end{figure}
The Scharber plot is based upon detailed balance, which we have shown fails
for the acceptor treated in this article.  In order to make semi-quantitative
educated guesses about how our candidate molecules might perform in real 
solar cells, we propose a simple model which we call the finite-width (fw) 
\marginpar{\color{blue} fw}
model. This model, while remaining computationally simple, tries to capture 
the fundamental physics that not every photon whose energy exceeds the band 
gap is going to be absorbed.  To this end, we assume that both the ACC and 
the DON
absorb photons in the wavelength range $(\lambda_g , \lambda_g + \Delta \lambda)$
where the gap wavelength $\lambda_g$ differs for the ACC and the DON but we will
insist on the same $\Delta \lambda$ cut-off for both the ACC and the DON.  Then
there are three different cases for any given photon wavelength: (i) it is neither
within the absorption range of the DON nor of the ACC, (ii) it is within the
absorption range of the DON or of the ACC but not both, or (iii) it is within
the absorption range of both the DON and of the ACC and so has a 50/50 chance
of being absorbed by each.  We will also assume that $\lambda_g^{ACC} > \lambda_g^{DON}$.
This leads to a fairly simple formula for the short-circuit current in our fw model,
\begin{equation}
  J_{fw} = \frac{\min \left( \Delta \lambda , \lambda_g^{ACC} - \lambda_g^{DON} \right)
  + \Delta \lambda}{2} \left( \lambda_g^{DON} F(\lambda_g^{DON})
  + \lambda_g^{ACC} F(\lambda_g^{ACC}) \right) \, .
  \label{eq:fw}
\end{equation}
{\bf Figure~\ref{fig:FiniteWidthJsc}} shows that this greatly overly simplified model
works well for predicting $J_{sc}$.  Hence we propose a modified Scharber model
in which the db approximation is replaced by our fw approximation.  Other parameters 
will be as above.  This then provides a well-defined way to estimate solar cell 
performance via a modified Scharber plot based upon our fw model and using only 
data that we may can calculate.

\subsection{Comparison of Chemical Properties}

Our ultimate interest is in predicting physical properties of importance
for DON:ACC solar cells. 
The ACC molecules considered in this paper have
the A-D-A form (see Fig.~\ref{fig:Fig01})
first adapted as a strategy for making non-fullerene acceptors in
2015 \cite{LWZ+15,LZB+15}.  In the previous subsection, we validated our 
TMC. 
The purpose of this subsection is to look at our candidate ACC molecules 
as a chemist might do.  

\begin{figure}
\begin{center}
\includegraphics[width=0.8\textwidth]{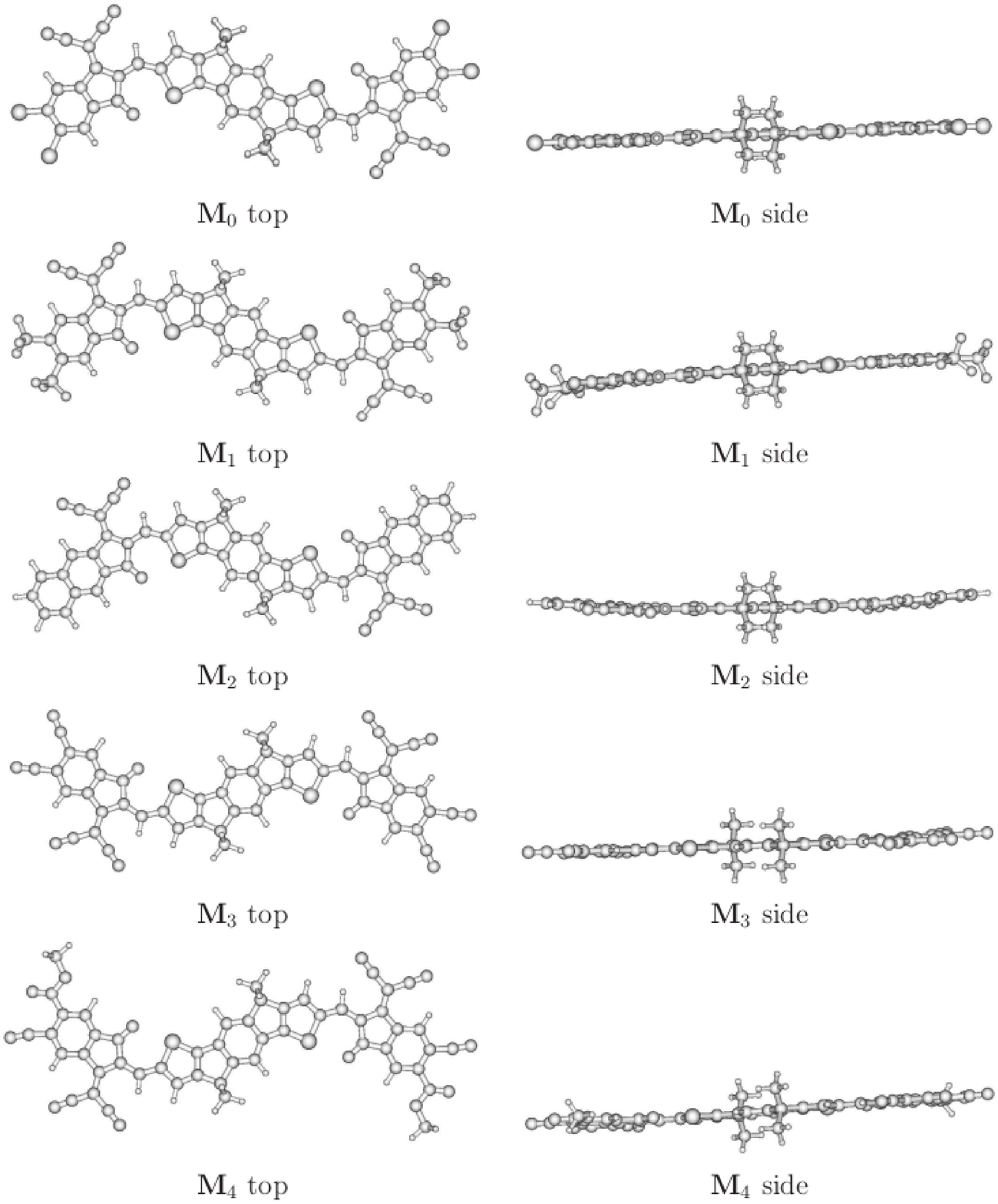}
\end{center}
\caption{
HSEH1PBE/6-311G(d,p)/PCM(chloroform) optimized geometries
as visualized with {\sc Molden} \cite{molden}.
\label{fig:geom}
}
\end{figure}
The first thing that would like to consider are the optimized geometries
shown in {\bf Fig.~\ref{fig:geom}}.  They are essentially flat, with some
minor curvature and/or twisting seen in the side view, but with methyl
side groups sticking up out of the plane of the central D section.
We do not study the effect of varying the
side groups as they are mainly a way of controlling morphology.  
A crystal structure for an A-D-A ACC \cite{QCZ+18} shows that the side 
groups prevent $\pi$ stacking of the D group.  Instead the A groups 
$\pi$ stack, forming a (presumably) conducting pile of aromatic rings.
This is illustrated schematically in Fig.~\ref{fig:stacking}.
Real organic solar cells are not crystalline but locally-crystalline
aggregation is likely.  Such locally-crystallinity has its pros and cons.  
On the one hand, the existance of crystal bands may help the charge-transfer
(CT) state to evolve more quickly into fully charge-separated (CS) states
\marginpar{\color{blue} CT, CS}
by favoring band-like conductivity over hopping-like conductivity.  Rapid
charge separation also reduces the possiblity for electron-hole recombination
which is detrimental to solar cell efficiency.  On the other hand, too 
much crystallinity may interfere with demixing of the ACC and DON, 
reducing the effective surface area of the ACC:DON
heterojunction and possibly making the solar cell too mechanically rigid for
some applications.  This is why bulky side groups are typically
added to be able to control the amount of aggregation so that it is neither
too much nor too little.


\begin{table}

\begin{center}
\begin{tabular}{cccccc}
\hline \hline
Compound & CV HOMO energy & CV LUMO energy & $\mu$ & $\eta$ & $\omega$ \\ 
\hline 
\multicolumn{6}{c}{Expt.}\\
IDIC(tetrahexyl)  \cite{LHZ+16}
                  & -5.69 eV  & -3.91 eV  & -4.80 eV & 1.78 eV & 6.47 eV \\
IDIC(tetrahexyl)-4Cl \cite{LMS+19} 
                  & -5.81 eV  & -4.01 eV  & -4.91 eV & 1.80 eV & 6.70 eV \\
\multicolumn{6}{c}{$\Delta$SCF}\\
{\bf M$_0$} & -5.77 eV  & -3.97 eV  & -4.87 eV  & 1.80 eV  & 6.59 eV \\
{\bf M$_1$} & -5.84 eV  & -4.07 eV  & -4.96 eV  & 1.77 eV  & 6.94 eV \\
{\bf M$_2$} & -5.70 eV  & -3.93 eV  & -4.82 eV  & 1.77 eV  & 6.55 eV \\
{\bf M$_3$} & -5.89 eV  & -4.18 eV  & -5.04 eV  & 1.71 eV  & 7.41 eV \\
{\bf M$_4$} & -5.84 eV  & -4.10 eV  & -4.97 eV  & 1.74 eV  & 7.10 eV \\
\multicolumn{6}{c}{``Koopmans''}\\
{\bf M$_0$} & -5.75 eV & -4.00 eV & -4.88 eV & 1.75 eV & 6.79 eV \\
{\bf M$_1$} & -5.82 eV & -4.10 eV & -4.96 eV & 1.72 eV & 7.15 eV \\
{\bf M$_2$} & -5.68 eV & -3.95 eV & -4.82 eV & 1.73 eV & 6.70 eV \\
{\bf M$_3$} & -5.87 eV & -4.21 eV & -5.04 eV & 1.66 eV & 7.65 eV \\
{\bf M$_4$} & -5.82 eV & -4.13 eV & -4.98 eV & 1.69 eV & 7.32 eV \\
\hline \hline
\end{tabular}
\end{center}

\caption{
CV FMO energies and CDFT reactivity indices calculated at the 
HSEH1PBE/6-311G(d,p)/PCM(acetonitrile) level.  
\label{tab:CDFT}}

\end{table}

The ACC molecules {\bf M}$_0$, {\bf M}$_1$, {\bf M}$_2$, {\bf M}$_3$, 
and {\bf M}$_4$ differ by modifications to the terminal A units in the 
A-D-A structure.  Let us try to understand how these modifications
affect the CV HOMO and CV LUMO energies, and the CV HOMO-LUMO gap in
terms of chemical ideas.
Many modern-day chemical concepts have only been given precise definitions
relatively recently through the advent of conceptual DFT (CDFT, 
\marginpar{\color{blue} CDFT}
see Ref.~\cite{L22} for a recent review), through the study of electron-number
and density derivatives of molecular energies.  In particular, CDFT helps to 
clarify the earlier
ideas of FMO theory (see Refs.~\cite{F76,A07}).
It should probably be pointed out before going much further that chemical
reactivity depends upon many factors.  Reactions may be under steric control,
charge control, or orbital control (as well as being affected by solvent
effects).  Pearson's well-known hard-soft acid-base (HSAB) theory concerns
\marginpar{\color{blue} HSAB}
charge- versus orbital-control. For present purposes, it is useful to look
at two density derivatives and their finite difference approximations.  
These are the chemical potential
\begin{equation}
  \mu = \left( \frac{\partial E}{\partial N} \right)_v 
  = \frac{E_{\mbox{CV HOMO}} + E_{\mbox{CV LUMO}}}{2} = -\chi \, ,
  \label{eq:results.01}
\end{equation}
which is just the negative of the Mullikan electronegativity ($\chi=-\mu$),
and the hardness
\begin{equation}
  \eta = \left(  \frac{\partial^2 E}{\partial N^2} \right)_v 
  = E_{\mbox{CV LUMO}}-E_{\mbox{CV HOMO}} \, .
  \label{eq:results.02}
\end{equation}
Note that we have adopted the more modern definition of hardness which
drops division by a factor of two.  Note also that we have used the 
IP = $-E_{\mbox{CV HOMO}}$ and EA = $-E_{\mbox{CV LUMO}}$ in our 
finite difference approximation which come from the
$\Delta$SCF calculations using the PCM implicit solvent model for
molecules dissolved in acetonitrile.  Interestingly our calculated
CV HOMO-LUMO gap in acetonitrile is now closer to the 
experimentally-measured quantity in acetonitrile
than was the case in Table~\ref{tab:FMOCl} where calculations
were carried out with chloroform as the solvent, thus further confirming
our choice of TMC.
Of particular importance for us is that these reactivity indices provide
information about how energies change when the number of electrons change.
In particular, placing the molecule in a constant chemical potential bath
results in the spontaneous transfer of $-\mu/\eta$ electrons.  This 
allows the electrophilicity index to be defined \cite{PVL99},
\begin{equation}
  \omega = \frac{\mu^2}{2 \eta} \, ,
   \label{eq:results.03}
\end{equation}
which is exactly the energy lowering due to charge transfer.
Numbers are given for compounds in this paper in {\bf Table~\ref{tab:CDFT}}.
{\em The columns marked $E_{\mbox{CV HOMO}}$ and $E_{\mbox{CV LUMO}}$ constitute
predictions to be confirmed by cyclic voltammetry in acetonitrile.}
The second part of the table (marked ``Koopmans'') goes beyond the finite 
difference approximation and uses the FMO approximation to calculate
the reactivity indices from directly DFT orbital energies.  This second
FMO approach is more approximate than the initial finite difference
approach.

\begin{figure}
\begin{center}
\includegraphics[width=0.6\textwidth]{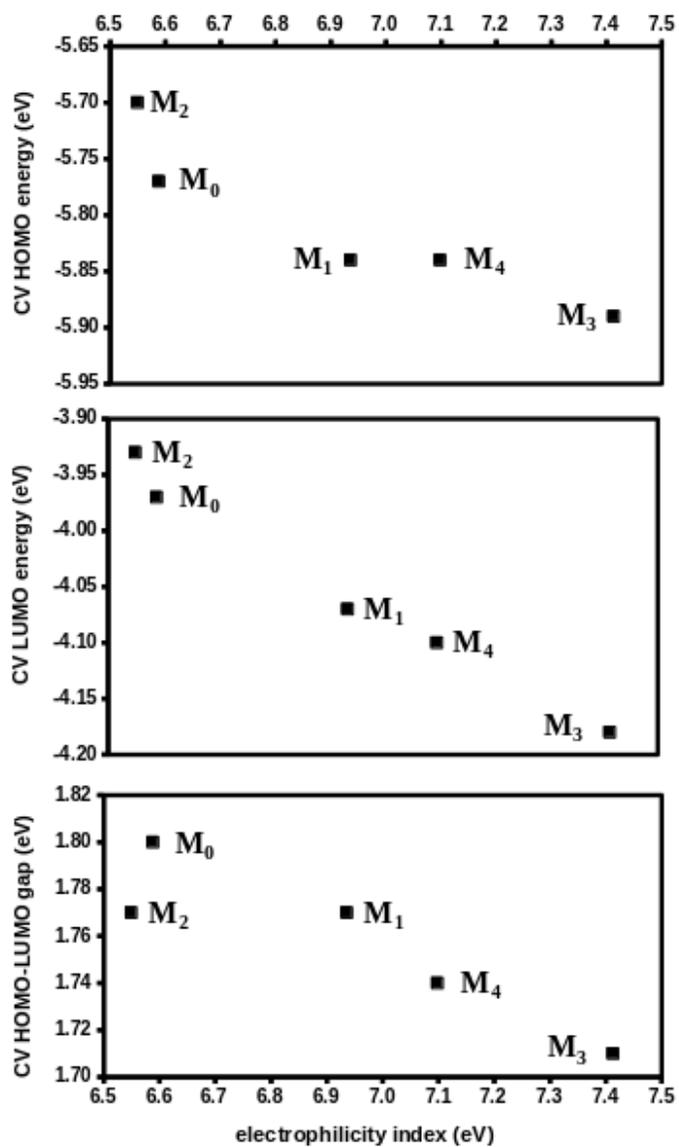}
\end{center}
\caption{
Correlation of CV FMO energies and the CV gap with the electrophilicity
index calculated from the CV FMO energies.
\label{fig:CDFT}
}
\end{figure}
The electrophilicity index allows us to organize our analysis of
key quantities as in {\bf Fig.~\ref{fig:CDFT}}.  FMO theory
predicts (see Fig.2 Supplementary information) that a good acceptor will lower the
energies of both the HOMO and the LUMO, so that both should decrease
as the electrophilicity index increases and this is indeed what is
seen in Fig.~\ref{fig:CDFT}.  However the CV LUMO energy is much
better linearly correlated with the electrophilicity index than is the
CV HOMO energy, as might be expected given that the electrophilicity
index is specifically designed to describe the stabilization resulting 
from donation of a pair of electrons into the LUMO.  It is also apparent
that the gap tends to decrease as the electophilicity index increases,
consistent with the appearance of the gap ($\eta$) in the denominator
of the definition of the electophilicity index [Eq.~(\ref{eq:results.03})].
However this is only a rough tendency and does not prevent {\bf M$_3$} 
from having a smaller gap than {\bf M$_4$}. The exact order of the
acceptor strength depends upon how it is defined.  If the definition is
based upon the CV HOMO energy, then we have the following order of
\begin{equation}
  \mbox{Acceptor Strength: {\bf M$_3$} $>$ {\bf M$_4$} $\approx$ 
  {\bf M$_1$} $>$ {\bf M$_0$} $>$ {\bf M$_2$}} \, .
  \label{eq:results.04}
\end{equation}
However basing our definition on the electrophilicity index (or the 
CV LUMO energy) gives the more definite order of
\begin{equation}
  \mbox{Acceptor Strength: {\bf M$_3$} $>$ {\bf M$_4$} $>$ 
  {\bf M$_1$} $>$ {\bf M$_0$} $>$ {\bf M$_2$}} \, .
  \label{eq:results.05}
\end{equation}


\begin{table}

\begin{center}
\begin{tabular}{cccc}
\hline \hline
Group & Hammett $\sigma_m$ & Hammett $\sigma_p$ & Remya \& Suresh $\Delta V_C$ \\
\hline 
CN           & 0.56 & 0.66 & 18.0 \\
CF$_3$       & 0.43 & 0.54 & 13.7 \\
COOMe        & 0.37 & 0.45 &  8.2 \\
Cl           & 0.37 & 0.23 &  7.0 \\
\hline \hline
\end{tabular}
\end{center}

\caption{
Some measures of the electron accepting capacity of chemical groups.
The Hammett $\sigma_m$ and $\sigma_p$ values have been taken from 
the aqueous values in Table~X of Ref.~\cite{HLT91}.  
The Remya \& Suresh $\Delta V_C$ have been taken from Table~1 of 
Ref.~\cite{RS16}.
\label{tab:Hammett}}

\end{table}

But does this make sense chemically?  That is, {\bf M$_0$}, {\bf M$_1$},
{\bf M$_2$}, {\bf M$_3$}, and {\bf M$_4$} differ by substitution of one
functional group for another.  Chemists can more easily recognize whether 
a functional group is electron donating or electron accepting than they
can quantify that donating and accepting power.  However the traditional
way to do this in physical organic chemistry \cite{J79} is to use the
Hammett $\sigma$ which is the difference in the $pK_a$ of benzoic acid
and benzoic acid substituted in the {\em meta} or {\em para} positions.
Some relevant Hammett $\sigma_m$ and $\sigma_p$ are tabulated in 
{\bf Table~\ref{tab:Hammett}}.  The larger these $\sigma$, the stronger
the electron acceptor.  Remya and Suresh have recently developed another
measure of donating and accepting power, but based upon differences of
the molecular electrostatic potential at the nucleus of the carbon atom
in the {\em para} position in substituted benzenes \cite{RS16}.
Their values of $\Delta V_C$ is also tabulated in Table~\ref{tab:Hammett}.
If we assume that any of these values are additive, then we arrive at
the following orders of accepting power for various combinations of
accepting groups:
\begin{eqnarray}
  2 \sigma_m(\mbox{CN}) = 1.12
  & > & \sigma_m(\mbox{CN}) + \sigma_m(\mbox{COOMe}) = 0.93
  > 2 \sigma_m(\mbox{CF$_3$}) = 0.86
  > 2 \sigma_m(\mbox{Cl}) = 0.74
  \nonumber \\
   2 \sigma_p(\mbox{CN})  = 1.32
  & > & \sigma_p(\mbox{CN}) + \sigma_p(\mbox{COOMe}) = 1.11
   \approx 2 \sigma_p(\mbox{CF$_3$}) = 1.08
  > 2 \sigma_m(\mbox{Cl}) = 0.46
  \nonumber \\
   2 \Delta V_C(\mbox{CN}) = 36.0
  & > & 2 \Delta V_C(\mbox{CF$_3$}) = 27.4
  \approx \Delta V_C(\mbox{CN}) + \Delta V_C(\mbox{COOMe}) = 26.2
  > 2 \Delta V_C(\mbox{Cl}) = 14.0 \, . \nonumber \\
  \label{eq:results.06}
\end{eqnarray}
This tells us nothing about the end group used in making {\bf M$_2$} 
but otherwise confirms Eq.~(\ref{eq:results.04}) [Eq.~(\ref{eq:results.05}) 
in the case of $\sigma_m$].  Hence our results are indeed consistent with 
those of traditional chemical theory.

\subsection{Prediction of Physical Properties}

We turn now to the prediction of physical properties for our candidate
molecules.  Of course, the CV FMO energies given in Table~\ref{tab:CDFT},
and extensively discussed in terms of their chemical meaning,
are already one prediction that may be confirmed by cyclic voltammetry
in acetonitrile.  Other important properties are the absorption spectrum
in chloroform and predictions of solar cell performance.

\begin{figure}
\begin{center}
\includegraphics[width=0.6\textwidth]{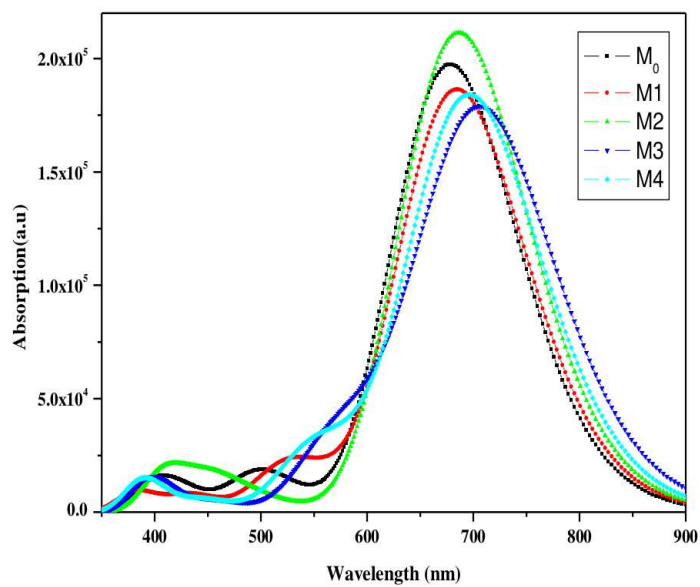}
\end{center}
\caption{
Absorption spectra calculated at the TD-HSEH1PBE/6-311G(d,p)/PCM(chloroform)
level.  Note that Epsilon ($\epsilon$) is the molar extinction coefficient.
\label{fig:Fig05}
}
\end{figure}

\begin{table}

\begin{center}
\begin{tabular}{ccccc}
\hline \hline
Molecules & $\lambda_{\mbox{max}}$ & $E_{\mbox{opt}}$ & $f$ & $R_{g,e}$ \\
\hline 
{\bf M$_0$} & 678 nm & 1.83 eV & 2.73 & 4.13 {\AA} \\
{\bf M$_1$} & 684 nm & 1.81 eV & 2.57 & 4.03 {\AA} \\
{\bf M$_2$} & 686 nm & 1.81 eV & 2.92 & 4.29 {\AA} \\
{\bf M$_3$} & 708 nm & 1.75 eV & 2.44 & 3.99 {\AA} \\
{\bf M$_4$} & 696 nm & 1.78 eV & 2.53 & 4.03 {\AA} \\
\hline \hline
\end{tabular}
\end{center}

\caption{Calculated absorption maxima ($\lambda_{\mbox{max}} = 
E_{\mbox{opt}}/hc$, where ``opt'' stands for ``optical'') 
and oscillator strengths ($f$) calculated
at the TD-HSEH1PBE/6-311G(d,p)/PCM(chloroform) level.  All transitions
are essentially pure singlet HOMO $\rightarrow$ LUMO.
The quantity $R_{g,e}$ is a measure of charge separation distance as
calculated from the transition dipole moment [see Eq.~(\ref{eq:results.09})].
\label{tab:tab3}}

\end{table}

\begin{figure}
\begin{center}
\includegraphics[width=0.8\textwidth]{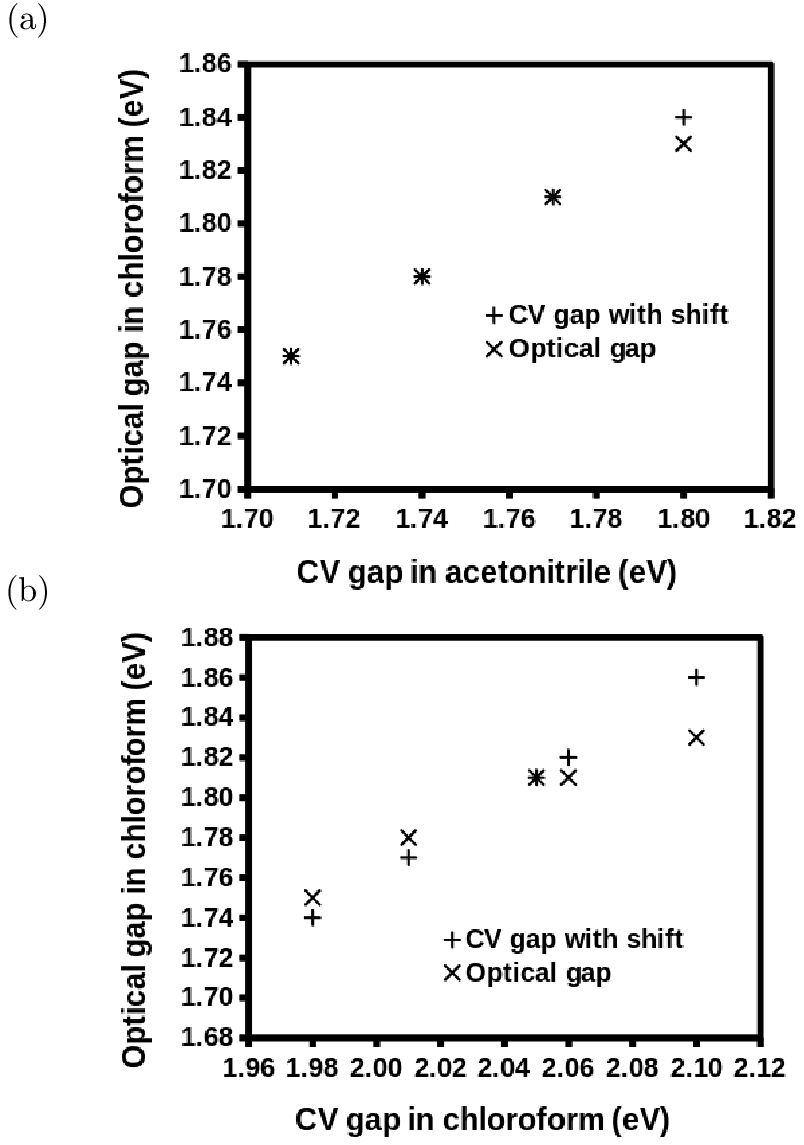}
\end{center}
\caption{
Correlation between the CV gap calculated in (a) acetonitrile and in (b)
chloroform and the energy of the absorption maximum (optical gap) 
calculated in chloroform.  When the CV gap is shifted down by an exciton 
binding energy of -0.04 eV (CV gap in acetonitrile) or 0.24 eV (CV gap 
in chloroform), then the shifted CV gap nearly co\"{\i}ncides with the 
optical gap.
\label{fig:gap_max_correl}
}
\end{figure}

\begin{table}

\begin{center}
\begin{tabular}{cccccc}
\hline \hline
Compounds & IP (eV) & EA (eV) & $E_{\mbox{fund}}$ (eV) & $E_{\mbox{opt}}$ (eV)
& $E_b$ (eV) \\
\hline 
{\bf M$_0$} & -5.99 & -3.89 & 2.10 & 1.82 & 0.28 \\
{\bf M$_1$} & -6.06 & -4.00 & 2.06 & 1.81 & 0.25 \\
{\bf M$_2$} & -5.85 & -3.80 & 2.05 & 1.81 & 0.24 \\
{\bf M$_3$} & -6.14 & -4.14 & 1.98 & 1.75 & 0.23 \\
{\bf M$_4$} & -6.06 & -4.05 & 2.01 & 1.78 & 0.23 \\
\hline \hline
\end{tabular}
\end{center}

\caption{
Calculation results of IP, EA, fundamental gap, first singlet excitation 
energies, and exciton binding energies of the reference and investigated 
acceptor molecules.
\label{tab:tab8}}

\end{table}

Calculated absorption spectra are shown in {\bf Fig.~\ref{fig:Fig05}}.
They are remarkably similar but there are small shifts in the position
of the main peak ({\bf Table~\ref{tab:tab3}}).  The calculations show that
the main peak is the 
singlet HOMO $\rightarrow$
LUMO transition.  As such, we might expect a good correlation between the
energy of the main peak of the chloroform absorption spectrum with the
CV gap in acetonitrile.  This is exactly what is seen in 
{\bf Fig.~\ref{fig:gap_max_correl}}(a) with the absorption maximum
being almost exactly 0.04 eV higher than the CV HOMO-LUMO gap.
Upon reflection, the near equivalence of the optical and CV HOMO-LUMO gaps 
is a little surprising as the shift has the wrong sign.  The exciton 
binding energy $E_b$ for a molecule {\bf M} 
\marginpar{\color{blue} $E_b$}
is $\Delta E$ for the reaction,
\begin{equation}
  \mbox{{\bf M$^*$} + {\bf M$^*$}} \rightarrow \mbox{{\bf M$^+$}}
  + \mbox{{\bf M$^-$}}  \, .
  \label{eq:results.07}
\end{equation}
In the Hartree-Fock two orbital two electron model (TOTEM) with
\marginpar{\color{blue} TOTEM}
frozen orbitals \cite{C95}, 
\begin{equation}
  E_b = \mbox{CV LUMO} - \mbox{CV HOMO} - [LL \vert HH] + [HL \vert LH]
  \, ,
  \label{eq:results.08}
\end{equation}
where the HOMO-LUMO coulomb integral $[LL \vert HH] > 0 $ is greater than
the HOMO-LUMO exchange integral $[HL \vert LH] > 0$.  Hence the exciton
binding energy $E_b$ is expected to be {\em larger} than the fundamental 
gap ($\mbox{CV LUMO} - \mbox{CV HOMO}$). In fact, it is {\em when the optical
and fundamental gaps are calculated in the same solvent}.  Table~\ref{tab:tab8}
shows the CV FMO energies recalculated in chloroform, instead of acetonitrile.
This leads to the plot shown in Fig.~\ref{fig:gap_max_correl}(b) where it is
seen that the exciton binding energy is indeed larger than the CV HOMO-LUMO 
gap.  Comparing the optical gap calculated in chloroform with the CV HOMO-LUMO
gap calculated in the more polar acetonitrile lead to a fortuitous, but 
misleading, co\"{\i}ncidence of the two gaps, which has been corrected
when the same solvent is used throughout.

\begin{figure}
\begin{center}
\includegraphics[width=0.6\textwidth]{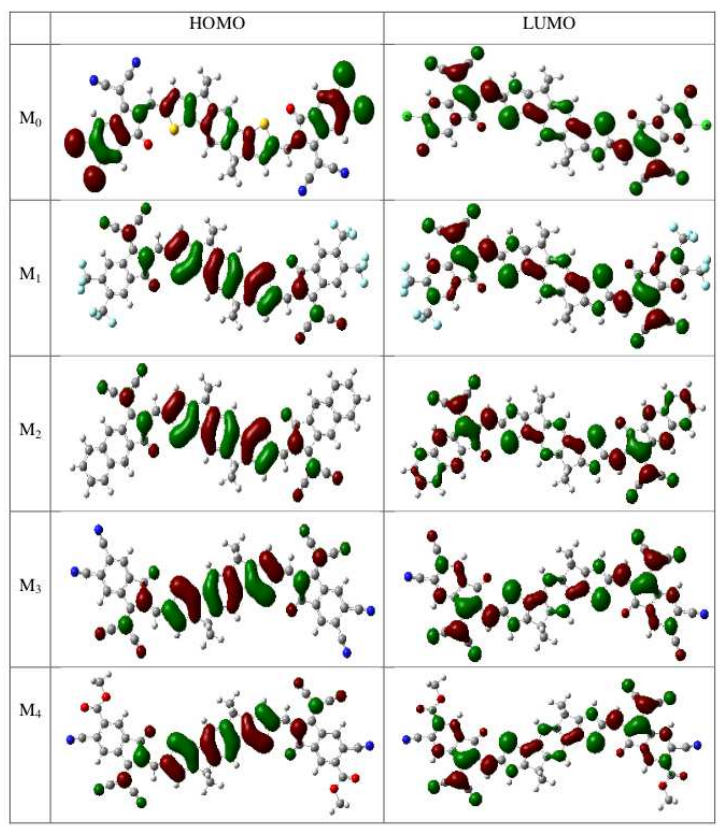}
\end{center}
\caption{
FMO isosurfaces calculated at the HSEH1PBE/6-311G(d,p)/PCM(chloroform) level.
\label{fig:Fig04}
}
\end{figure}
\begin{figure}
\begin{center}
\includegraphics[width=0.6\textwidth]{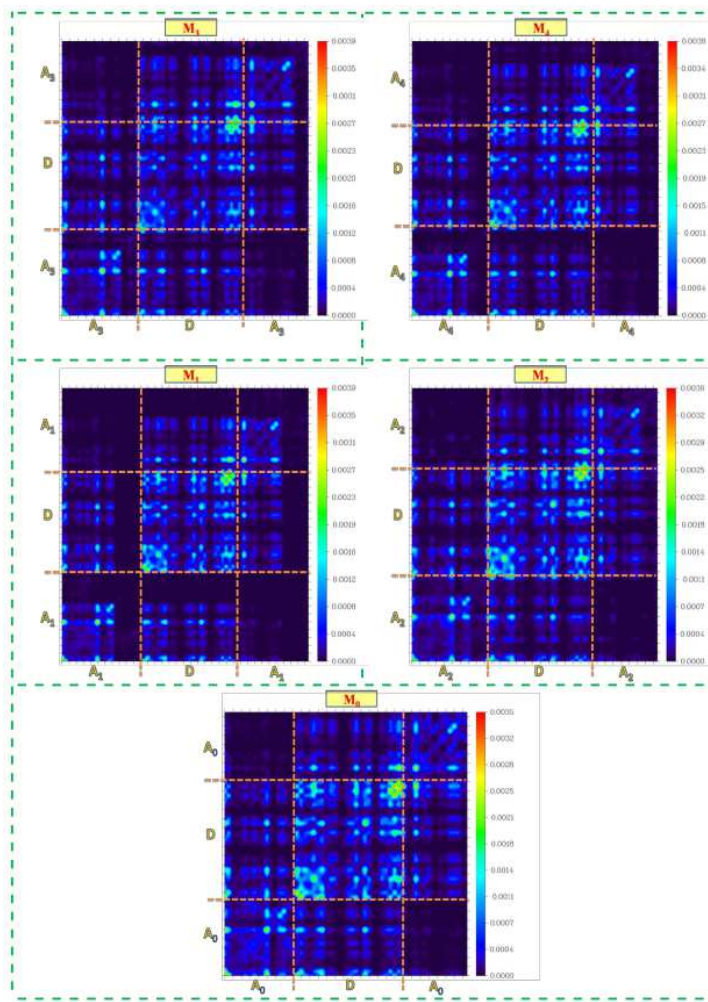}
\end{center}
\caption{
\label{fig:Fig07}
Heat maps of the $S_1$ excitation transition density matrices (TDMs)
calculated at the TD-HSEH1PBE/6-311G(d,p)/PCM(chloroform) level using
{\sc Gaussian} \cite{g09} and {\sc Multiwfn} \cite{LC12,Multiwfn}
where the charge-transfer numbers $\Omega_{A,B}$ have been calculated
using Way 4 as described on page 211 of Ref.~\cite{Multiwfn}.
Each pixel of the heat map is thus characterized by two atoms (A and B) 
and is symmetrized (the first equation at the top of page 211 of 
Ref.~\cite{Multiwfn}).  As is usual, hydrogen atoms have been omitted 
from the figure in order to keep the size and complexity of the figure 
manageable.  Note that no attempt has been made to insure that the atom 
numbering used in making the heat maps reflects the underlying symmetry of the
molecule.
}
\end{figure}

{\bf Figure~\ref{fig:Fig04}} shows the FMO isosurfaces for the 
the A-D-A molecules studied in this paper.  It is qualitatively clear
that HOMO density on D is being transferred to the LUMO density on A,
making them charge-transfer excitations.  Such excitations create 
electron-hole pairs whose dissociation leads to, but whose recombination
prevents, charge transport.  Characterizing these charge transfer
excitations is not as obvious as might first be assumed because
they may represent states that separate charge in equal but opposite
directions so that, on average, no real transport of charge 
takes place \cite{DCJ+18}.  A way around this dilema is to calculate and 
analyze the TDM 
\begin{equation}
  \gamma^{(0,I)}(1,2) = \langle \Psi_0 \vert \hat{\psi}^\dagger(2) 
  \hat{\psi}(1) \vert \Psi_I \rangle \, ,
  \label{eq:results.08.1}
\end{equation} 
for each transition $0 \rightarrow I$ condensed to atoms 
\begin{equation}
  \Omega_{A,B} = \int_A d1 \int_B d2 \, \vert \gamma^{(0,I)}(1,2) \vert^2
  \, ,
  \label{eq:results.08.2}
\end{equation}
where the first integral is over the electron coordinates associated with 
atom A and the second integral is over the hole coordinates associated with 
atom B.  The quantity $\hat{\psi}(i)$ is the field operator for the position 
of electron $i$.  The charge-transfer number for an electron-hole excitation
represents the overlap of the part of the hole on atom B with the part of
the hole on atom A and so tells how much of an electron has been excited
from atom A to atom B.  Of course, Eq.~(\ref{eq:results.08.2}) is incomplete 
until a way is specified to associate coordinates with atoms.  As there is
no unique way to do this, 
the exact evaluation of the charge-transfer 
numbers $\Omega_{A,B}$ depends upon particular choices of population analysis 
scheme \cite{PL12,PBWD14,PWD14,P20,Multiwfn}.  It may also be symmetrized \cite{Multiwfn}
in which case it is the average of how much of an electron has been excited
from A to B and how much has been excited from B to A.
Calculated heat maps of $S_1$ TDMs are shown in {\bf Fig.~\ref{fig:Fig07}}.  
As should be expected from the FMO maps (Fig.~\ref{fig:Fig04}), most
of the HOMO $\rightarrow$ LUMO excitation in these A-D-A molecules is
from localized on D with a lesser extent on A.  In this sense, D acts
as the chromophore whose excitation is to be transferred to the A ends
of the molecule.  This charge transfer, already seen with the FMOs, is
seen again by the presence of blue in the (D,A) and (A,D) blocks of the TDMs.  

Neither the pictures of FMOs nor the TDM heat maps provide quantitative
information about the electron-hole charge transfer.  While many 
approaches have been proposed to characterize the degree of electron-hole
separation, we will take a simple approach here by focusing
on the oscillator strengths in Table~\ref{tab:tab3} which are seen
to be between about 2.5 and 3.0.  Such large oscillator strengths are 
typical of charge transfer excitations.  In fact, the transition 
dipole moment $\mu_{g,e} = e R_{g,e}$ for the excitation with energy 
$\hbar \omega$ may be calculated from the exact oscillator strength expression,
\begin{equation}
  f = \frac{2}{3} \frac{\omega m_e}{\hbar} e^2 R_{g,e}^2 \, .
  \label{eq:results.09}
\end{equation}
Values of $R_{g,e}$ (also shown in Table~\ref{tab:tab3}) provide 
a simple measure of electron-hole separation.  We see that they
are on the order of the size of one A unit or less consistent with
physical expectations.  The values of $R_{g,e}$ make it clear that 
{\bf M}$_2$ is the best molecule when it comes to maximizing 
electron-hole separation.  It is also best in the sense of absorbing
the most light.

\begin{figure}
\begin{center}
\includegraphics[width=0.6\textwidth]{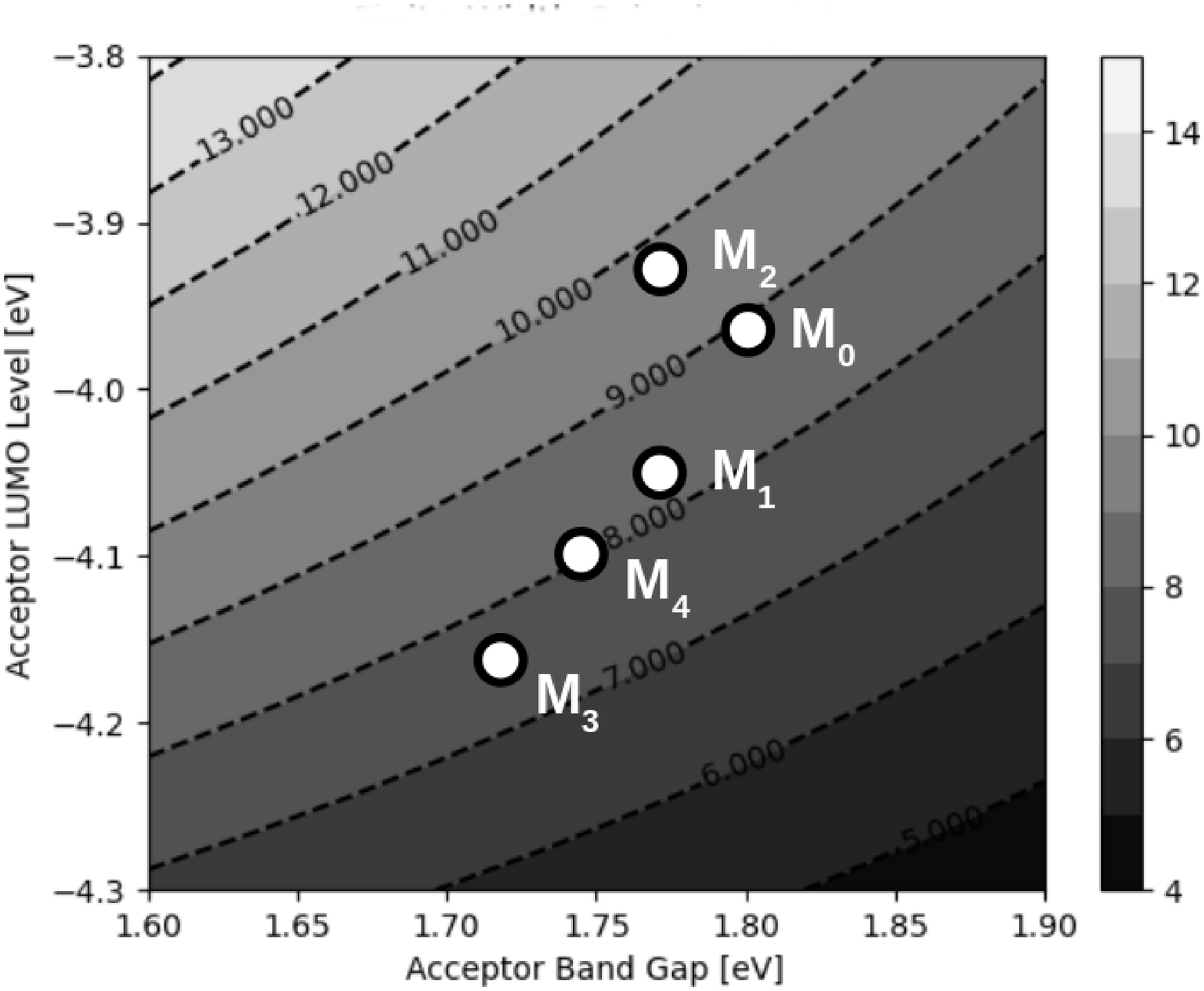}
\end{center}
\caption{
Scharber plot made using the finite width model.
\label{fig:fwScharber}
}
\end{figure}

\begin{table}

\begin{center}
\begin{tabular}{cccc}
\hline \hline
ACC & $eV_{oc}$ (eV) & $J_{sc}$ (mA/cm$^2$) & PCE \\
\hline 
\multicolumn{4}{c}{$\Delta SCF$}\\
{\bf M$_0$} & 0.79 eV  & 16.01 mA/cm$^2$ &  8.86 \% \\
{\bf M$_1$} & 0.69 eV  & 16.76 mA/cm$^2$ &  8.10 \% \\
{\bf M$_2$} & 0.83 eV  & 16.76 mA/cm$^2$ &  9.74 \% \\
{\bf M$_3$} & 0.58 eV  & 18.28 mA/cm$^2$ &  7.42 \% \\
{\bf M$_4$} & 0.66 eV  & 17.52 mA/cm$^2$ &  8.09 \% \\
\multicolumn{4}{c}{``Koopmans''}\\
{\bf M$_0$} & 0.76 eV  & 17.26 mA/cm$^2$ &  9.18 \% \\
{\bf M$_1$} & 0.66 eV  & 18.02 mA/cm$^2$ &  8.33 \% \\
{\bf M$_2$} & 0.81 eV  & 17.77 mA/cm$^2$ & 10.08 \% \\
{\bf M$_3$} & 0.55 eV  & 19.58 mA/cm$^2$ &  7.54 \% \\
{\bf M$_4$} & 0.63 eV  & 18.80 mA/cm$^2$ &  8.29 \% \\
\hline \hline
\end{tabular}
\end{center}

\caption{
Estimated PCEs for PM6:ACC cells obtained using Eq.~(\ref{eq:PCE}) with 
FF = 70\% , $P_s$ = 99.971 mW.cm$^2$, Eq.~(\ref{eq:Voc}) to calculate $V_{oc}$
with $\delta$ = 0.74 eV, and Eq.~(\ref{eq:fw}) to calculate $J_{sc} = J_{fw}$
with $\Delta \lambda$ = 200 nm.  The PM6 CV FMO energies are from 
Table~\ref{tab:PVCl} and the ACC CV FMO energies are from Table~\ref{tab:CDFT}.
\label{tab:PCE}}

\end{table}

We now wish to estimate solar cell performance with these candidate 
NFAs.  We cannot do this with the original 
Scharber plot method \cite{SMK+06,S16} (reviewed in the SI)
because that assumes that light absorption only occurs
in the DON, whereas we have used a detailed balance argument in 
Subsec.~\ref{sec:TMCA} to show that exciton formation due to light 
absorption must also be occurring in the ACC.  Instead, we will make
a Scharber plot for a PM6:ACC cell and our different candidate acceptors
using the fw model introduced in Subsec.~\ref{sec:TMCA}.
{\bf Figure~\ref{fig:fwScharber}} may be the first Scharber plot in the
literature whose axes are labeled by the ACC gap and LUMO levels, instead
of the usual DON quantities.  It shows that the main factor influencing
performance is the ACC LUMO level.  {\bf Table~\ref{tab:PCE}} provides 
specific numbers.  These differ a bit depending upon
whether the ACC FMO levels have been calculated using the $\Delta SCF$ CV
FMO levels or the DFT FMO levels, but general trends are independent of
which choice is made.  The candidate ACC which is predicted to give the
highest $J_{sc}$ is {\bf M$_3$} because it has the lowest HOMO-LUMO gap.
However this is outweighed by $eV_{oc}$, the order of PCE of the
different candidate ACCs being the same as the order of their $eV_{oc}$.
The largest $eV_{oc}$, and hence the largest PCE, is for {\bf M$_2$} 
because this molecule has the largest LUMO energy.  Hence the energy of
the photon exciting from the DON HOMO to the ACC LUMO is maximized and
so is the energy needed for efficient charge separation.  
Figure~\ref{fig:fwScharber} suggests that further progress could be made
by lowering the ACC band gap while maintaining a high ACC LUMO energy.

%
%

\section{Discussion and Conclusion}
\label{sec:conclude}

The advent of third generation bulk heterojunction OSCs is bringing 
OSC PCEs up to nearly 20\%, making it possible to envisage OSCs moving
out of niche markets to become an essential player in the evolving
world of energy technology.  OSCs use a DON:ACC architecture.  First
generation OSC design concentrated on P3HT:PC$_{61}$M-like OSCs where
the major issue was cell morphology.  The second generation mostly
concentrated on replacing P3HT with more optimal donor compounds.
The third and present generation consists of optimizing the ACC,
replacing the PC$_{61}$M fullerene derivative with NFAs.  These
NFAs typically consist themselves of a push-pull architecture involving
joined A and D units.  They are typically flat molecules
that stack well so as to form locally crystalline structures.  This
allows the rapid movement of charges away from the DON:ACC interface,
hence minimizing electron-hole recombination and increasing cell performance.
However too much crystallinity can also lead to undesirable OSC morphologies
which decrease OSC performance.  It has not been the purpose of this
article to examine either morphologies or aggregation.  Rather we begin
with the known A-D-A ACC IDIC-4Cl and replace the Cl end groups
with other end groups making the reasonable assumption that such small
changes will have little affect on morphology but will nevertheless
affect the OSC PCE.  It is to be emphasized that, while IDIC-4Cl does
not yield PCEs of more than about 10\%, the relatively simple A-D-A
structure makes it possible to understand its performance in quite
some detail.  In particular, small structural variations are expected
to lead to small variations in the corresponding PCE and we have
analyzed the effects of these these variations.

We began with an extensive validation in which we were able to confirm 
that the presence of side chains on IDIC-4Cl derivatives has little 
effect on core electronic properties.  The molecules are rigid and 
geometry optimizations match well against the single crystal structure
used for comparison purposes.  Calculated CV FMOs agreed well with experiment.
Recalculating using the same solvent as used in the experimental measurements
further improves agreement between calculation and experiment.
Absorption spectra calculated with TD-DFT were also in quite reasonable
agreement with experiment.  The DON:ACC OSC PCE was found to depend primarily 
on the $J_{sc}$ hich, somewhat surprisingly, does not simply 
increase as the ACC HOMO-LUMO gap increases.  This is counter to what was
found in second-generation OSCs where the short-circuit current generally
follows the db model and always grows as the DON HOMO-LUMO gap
increases.  The difficulty was traced back to the need to take into account
light absorption in both the DON and ACC materials.  We have modeled
this by introducing a fw approximation into Scharber's theory 
for estimating OSC PCE performance.  In Scharber's plots PCE increases 
with both the DON HOMO-LUMO gap and the DON LUMO energy.
In our fw theory, PCE increases with the ACC LUMO energy but
actually decreases as the ACC HOMO-LUMO gap grows and approaches the DON
HOMO-LUMO gap.  Our theory is very simple, though adequate for present 
purposes.   

The rough effects of electron donation and electron accepting groups on
the HOMO and LUMO energies and on the HOMO-LUMO gap is well established
in FMO theory.  Determining the relative strength of electron acceptors
is not always {\em a priori} obvious, especially when electron acceptors
are of similar strength.  We have used conceptual DFT
to calculate electrophilicity indices and showed that the CV FMO energies
and the CV HOMO-LUMO gap all tend to decrease as the electrophilicity
index increases.  These results were confirmed by showing that the tends
match very well both the classic theory of Hammett but also the newer
theory of Remya and Suresh.   Our finite-width model predicts that the
DON:ACC OSC PCE will be especially sensitive to the ACC LUMO energy, 
with the ACC having the highest LUMO energy giving the highest PCE.
We thus predict that only one of the four candidate ACC molecules 
will outperform the orginal ACC molecule and have understood why
the other candidate molecules should have lower PCEs.  This same candidate
ACC molecule also has the largest oscillator strength for the important
$^1$(HOMO,LUMO) charge-transfer transition and the charge transfer is
shown to be more delocalized than for the other molecules studied here.

\section*{Acknowledgments}

It is both appropriate and a pleasure to have this article appear in a 
volume honoring Carlo Adamo on the occasion of his 60th birthday.  Carlo
has made important contributions in many areas of methodology used in the
present paper, including the development of density-functional approximations,
implicit solvent models, and simulations of photoprocesses.

We would like to thank Pierre Girard, Denis Charapoff, 
S\'ebastien Morin, and for technical support in the context 
of the Grenoble {\em Centre d'Experimentation du Calcul Intensif en Chimie}
({\em CECIC}) computers used for the calculations reported here.  Mourad Chemek,
Tarek Mestiri, and Malak Hijazi are thanked for helpful discussions.

\section*{Author Contributions}

Author credit has been assigned using the CRediT contributor roles taxonomy system \cite{CRediT}.
\\
{\bf Walid Taouali}: Conceptualization, investigation, formal analysis, writing - original draft, visualisation.
\\
{\bf Kamel Alimi}: Writing - review \& editing.
\\
{\bf Asma Sindhoo Nangraj}: Visualisation, software.
\\
{\bf Mark E.\ Casida}: Supervision, resources, formal analysis, writing - original draft, visualisation, software.

\section*{Conflicts of Interest}

The authors declare no conflict of interest.

\section*{Supplementary Information}
\label{sec:SI}

\begin{enumerate}
 \item Scharber Plots
 \item Chemical Names
 \item Optimized Structures
 \item FMO theory diagram
 \item Spectral shift
\end{enumerate}

\singlespace

\end{document}


\maketitle

\tableofcontents

\section{Scharber Plots}

We wrote our own computer program to make ``Scharber plots''
(also known as ``Scharber diagrams'') \cite{S16}.  We review
the basic theory as we had to clarify certain points in the 
original description of the phenomenological theory \cite{SMK+06} 
in order to be able to write our own computer program.  Note that 
this allows us to apply the same basic ideas more widely than did 
Scharber should we wish to do so.  {\em Key assumptions 
are given in italics.}

Experimentalists who seek to characterize a new material for potential
use in a solar cell, generally illuminate the cell with AM 1.5 G light
(solar spectrum at ground level as seen through an air mass of 1.5 times 
that of the atmosphere) and power $P_s$.  They then make a $(V,J)$-plot 
whose general form is usually well described by the 
Schokley diode equation \cite{MMAC22}.  When the voltage $V=0$, then 
we obtain the short-circuit current density $J_{sc}$ (mA/cm$^2$).  When 
the current density $J=0$, then we obtain the open-circuit voltage 
$V_{oc}$ (V).  Inscribing a square inside the $(V,J)$-curve defines the 
point $(V_m,J_m)$.  Since the product $J_m V_m$ is also the maximum power 
that can be extracted from the solar cell, the power conversion efficiency 
(PCE) of the solar cell is,
\begin{equation}
  \eta = \frac{V_m J_m}{P_s} \, .
  \label{eq:theory.1}
\end{equation}
Shockley and Queisser showed for a simple solar cell that $\eta$ has a 
thermodynamic upper limit \cite{SQ61} which today is accepted to be 34\% 
at AM 1.5G.  Organic solar cells are significantly less efficient.

Another quantity that may be determined from a $(V,J)$-plot is the
fill factor (FF), given by
\begin{equation}
  \mbox{FF} = \frac{V_m J_m}{V_{oc} J_{sc}} \, .
  \label{eq:theory.2}
\end{equation}
According to Gr\"atzel \cite{G09}, ``The value of the fill factor reflects 
the extent of electrical (Ohmic) and electrochemical (overvoltage) losses.''
{\em In constructing the Scharber plot, we use a typical value of 
FF = 0.65.}  Combining Eqs.~(\ref{eq:theory.1}) and (\ref{eq:theory.2})
gives,
\begin{equation}
  \eta = \frac{J_{sc} V_{oc} \mbox{FF}}{P_s} \, ,
  \label{eq:theory.3}
\end{equation}

For good electrical contacts, the open-circuit voltage is almost
given by just the difference of the CV HOMO energy of the 
molecules making up the acceptor layer $\epsilon_{\mbox{HOMO}}^{\mbox{DON}}$
and of the CV LUMO energy $\epsilon_{\mbox{LUMO}}^{\mbox{ACC}}$
of the molecules making up the donor layer.  It can be shown that,
\begin{equation}
  e V_{oc} \leq \epsilon_{\mbox{LUMO}}^{\mbox{ACC}} -
    \epsilon_{\mbox{HOMO}}^{\mbox{DON}} \, ,
  \label{eq:theory.4}
\end{equation}
with near equality obtained for good (ohmic) contacts between the
solar cell and the electrodes.  Empirically for a number of polymer
heterojunction solar cells where the acceptor molecule is PCBM,
near equality is found --- that is,
\begin{equation}
  V_{oc} = \frac{1}{e} \left( \epsilon_{\mbox{LUMO}}^{\mbox{PCBM}} -  
  \epsilon_{\mbox{HOMO}}^{\mbox{DON}} \right) - 0.3 \mbox{ V}
  \, .
  \label{eq:theory.6}
\end{equation}
{\em This is a second assumption used in constructing the
Scharber plot.}  Equation~(\ref{eq:theory.6}) is ``Equation 1 for 
$V_{oc}$'' in Ref.~\cite{SMK+06}.  We note that the original Scharber 
plot used
\begin{equation}
  \epsilon_{\mbox{LUMO}}^{\mbox{PCBM}} = -4.3 \mbox{ eV} \, ,
  \label{eq:theory.7}
\end{equation}
{\em which is another assumption used in making the Scharber
plot.} (For comparison, recall that others \cite{B10,HL11} have 
reported that the PCBM CV LUMO is located at -3.83$\pm$0.15 eV \cite{B10}.)

\begin{figure}
\begin{center}
\includegraphics[width=0.8\textwidth]{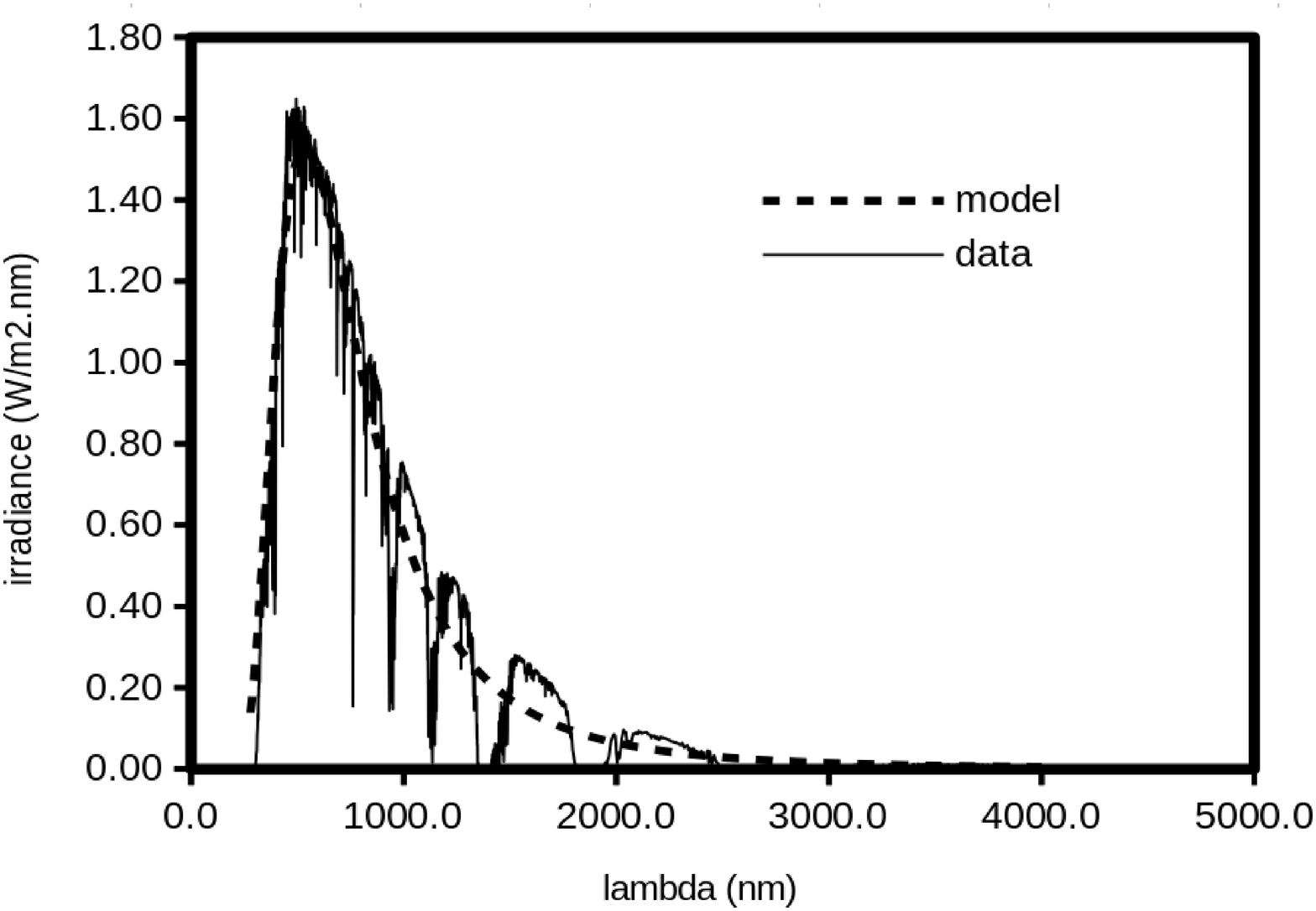}
\end{center}
\caption{
Comparison of the model spectral irradiance $F3(\lambda)$ obtained
from Eq.~(\ref{eq:theory.19}) via Eq.~(\ref{eq:theory.17})
with AM 1.5 G spectra irradiance data $F(\lambda)$ from Ref.~\cite{NREL}.
\label{fig:F3vsF}
}
\end{figure}

The penultimate quantity needed to be able to construct the Scharber plot
is $J_{oc}$.  {\em This is obtained by making the two detailed balance
assumptions}, namely: (i) every photon whose energy is greater than 
the fundamental gap,
\begin{equation}
  E_g = \epsilon^{\mbox{Donor}}_{\mbox{LUMO}} - \epsilon^{\mbox{Donor}}_{\mbox{HOMO}}
  \, ,
  \label{eq:theory.8}
\end{equation}
is absorbed and (ii) every absorbed photon creates an electron-hole
pair \cite{SQ61}.  {\em In constructing the Scharber plot, it
is assumed that the external quantum efficiency EQE = 65\% --- i.e.,  
that only 65\% of the photons are absorbed.}  We may then 
convert AM 1.5 Global data into a current density.  This is complicated 
by the fact that the data is reported 
as a spectral irradiance in units of watts/(m$^2$.nm).  To see how
to do the calculation, consider first monochromatic light of wave length
$\lambda$ hitting an area $A$ during a time $\Delta t$.  Let
$N$ be the number of photons hitting during this time.  Then the
current density is just,
\begin{equation}
  J = \frac{eN}{A\Delta t} \, .
  \label{eq:theory.9}
\end{equation}
The irradiance (${\cal I}$) is the energy per area per time.  So,
\begin{equation}
  {\cal I} = \frac{E}{A\Delta t} = \frac{Nh\nu}{A \Delta t}
  = \frac{Nhc}{A\Delta t} \frac{1}{\lambda} \, .
  \label{eq:theory.10}
\end{equation}
Hence,
\begin{equation}
	J = e \frac{{\cal I}\lambda}{hc} \, .
  \label{eq:theory.11}
\end{equation}
However AM 1.5G light is not monochromatic, so we must write that,
\begin{equation}
  J = \frac{e}{A \Delta t} \int_0^{\lambda_g}
  \, dN(\lambda) \, ,
  \label{eq:theory.12}
\end{equation}
where the gap wavelength is,
\begin{equation}
  \lambda_g = \frac{hc}{E_g} = \frac{1240 \mbox{ eV.nm}}{E_g} \, .
  \label{eq:theory.13}
\end{equation}
The spectral irradiance $F(\lambda)$ is defined as,
\begin{equation}
	F(\lambda) = \frac{d{\cal I}(\lambda)}{d\lambda} \, ,
  \label{eq:theory.14}
\end{equation}
so the irradiance for a small wavelength interval 
$(\lambda, \lambda+\Delta \lambda)$ is,
\begin{equation}
  F(\lambda) d\lambda = d{\cal I}(\lambda) = \frac{hc}{A\Delta t} 
  \frac{1}{\lambda} dN(\lambda) \, .
  \label{eq:theory.15}
\end{equation}
Then,
\begin{equation}
  dN(\lambda) = F(\lambda) \lambda \frac{A \Delta t}{hc} d\lambda \, ,
  \label{eq:theory.16}
\end{equation}
and,
\begin{equation}
  J = \frac{e}{hc} \int_0^{\lambda_g} F(\lambda) \lambda 
  \, d \lambda \, .
  \label{eq:theory.17}
\end{equation}
It remains only to calculate the constant $e/hc$ in appropriate
units:
\begin{equation}
  \frac{e}{hc} = \frac{
    (1.602 \times 10^{-19} \mbox{ C})(10^{+3} \mbox{ mA/(C/s)})
    }{
    (6.626 \times 10^{-34} \mbox{ J.s})(2.998 \times 10^8 \mbox{ m/s}) 
    (10^2 \mbox{ cm/m})^2 (10^9 \mbox{ nm/m})}
   = 8.0645 \times 10^{-5} \frac{\mbox{mA.m$^2$}}{\mbox{W.cm$^2$.nm}}
   \, .
  \label{eq:theory.18}
\end{equation}
We have carried out the integration using a spreadsheet and we have
found it to be very well approximated by the formula,
\begin{equation}
  J3(E_g) = (73.531 \mbox{ mA/cm$^2$}) e^{-(0.440)(E_g/\mbox{eV})^{1.8617}}
  \, ,
  \label{eq:theory.19}
\end{equation}
given as Eq.~(5) of Ref.~\cite{ARE+15}.  
Conversely, we may obtain an estimate $F3(\lambda)$ of the irradiance 
$F(\lambda)$ by taking the derivative of Eq.~(\ref{eq:theory.17}) to 
obtain,
\begin{equation}
  F3(\lambda) = \frac{hc}{e\lambda} \frac{d J3(hc/\lambda)}{d\lambda} \, .
  \label{eq:theory.19b}
\end{equation}
{\bf Figure~\ref{fig:F3vsF}} shows that $F3$ is a reasonable smoothed 
approximation to the standard AM 1.5 G irradiance data \cite{NREL}.

And the very last quantity that we need is the total power of the sun, $P_s$.
This can be obtained by directly integrating the AM 1.5G solar data,
\begin{equation}
  P_s = \int_0^\infty F(\lambda) \, d\lambda = 999.71 \mbox{ W/m$^2$} = 99.971 \mbox{ mW/cm$^2$} \, .
  \label{eq:theory.20}
\end{equation}
Combining all of the assumptions underlying the Scharber model into
a single formula then gives,
\begin{equation}
  \eta = \mbox{100\%} \times \frac{0.65 J3(E_g)}{99.971 \mbox{ mW.cm$^2$} } 
  \left[ \frac{1}{e} \left( \epsilon_{\mbox{LUMO}}^{\mbox{PCBM}} -
  \epsilon_{\mbox{HOMO}}^{\mbox{DON}} \right) - 0.3 \mbox{ V} \right] (0.65)
  \, .
  \label{eq:theory.21}
\end{equation}
We note that we have verified that our program gives Scharber plots in 
perfect agreement with plots in the original published work \cite{SMK+06}.
Moreover mastery of the Scharber formalism allows the same formalism to
be generalized to other contexts when, say, an exact FF is known experimentally
or the PCBM is replaced with a different ACC as is done in the present
work.

\section{Chemical Names}

It is common in this field to use abbreviations rather than long
chemical names. Moreover the same abbreviations are frequently reused
for the same molecules differing only by the nature of their side 
groups.  The full names and Lewis representations are given here
for molecules referred to only by abbreviations in the main text.

\begin{description}

\item[IDIC] The generic structure is
\begin{center}
\includegraphics[width=0.4\textwidth]{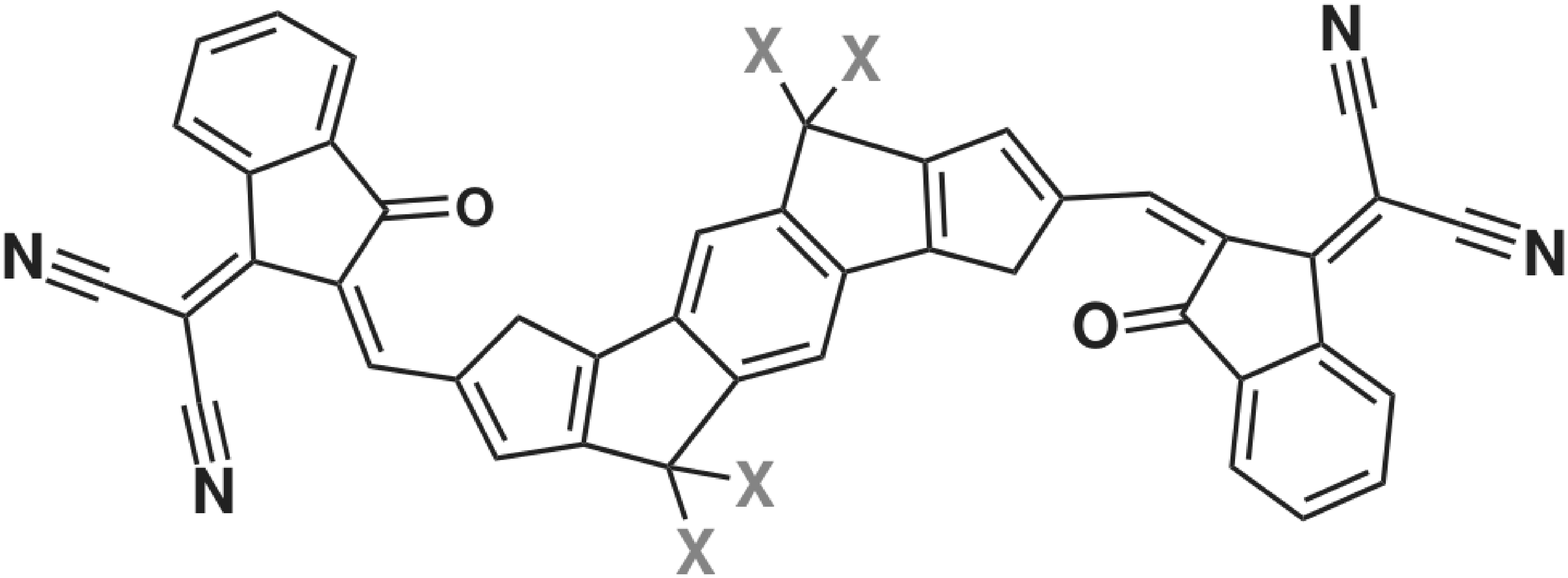}
\end{center}
and has the name
2,2'-((2Z,2'Z)-((4,4,9,9-X-4,9-dihydro-s-indaceno[1,2-b:5,6-b']dithiophene-2,7\\-diyl)bis(methanylylidene))bis(3-oxo-2,3-dihydro-1H-indene-2,1-diylidene))dimalononitrile
\\ where X stands for four equivalent groups.  When we want to be more specific,
then we will write IDIC(X).  For example, X should be replaced by 
``tetrahexyl'' in the orginal IDIC \cite{LHZ+16}, by ``tetraoctyl'' in 
Ref.~\cite{HZLZ+21}, by ``tetrakis(4-hexylphenyl)'' in the compound 
IDIC-4H of Ref.~\cite{QCZ+18}, and by ``tetramethyl'' for 
the simplified model compound for which calculations are actually done in 
this paper.

\item[IDIC-4Cl] The generic structure is 
\begin{center}
\includegraphics[width=0.4\textwidth]{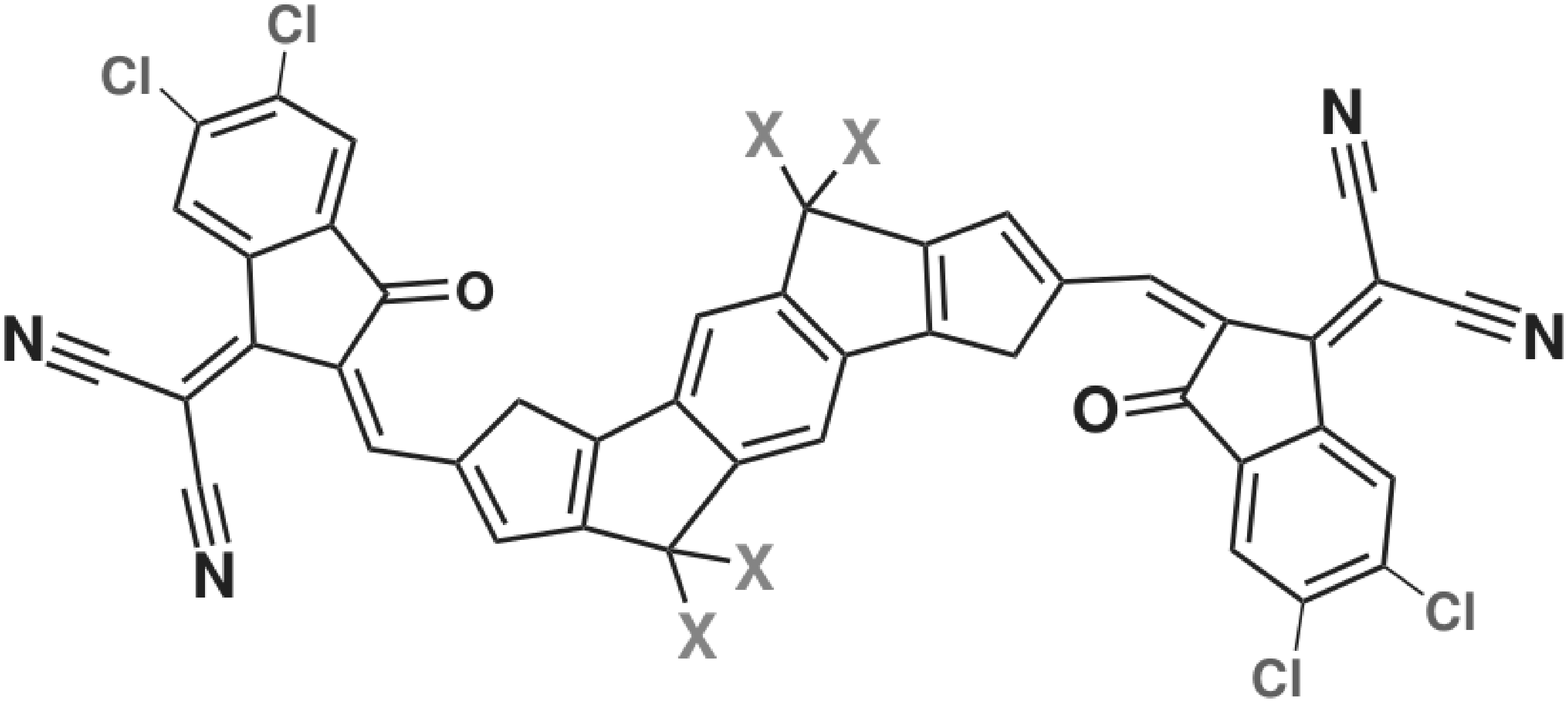}
\end{center}
and has the name
2,2'-((2Z,2'Z)-((4,4,9,9-X-4,9-dihydro-s-indaceno[1,2-b:5,6-b']dithiophen\\e-2,7-diyl)bis(methanylylidene))bis(5,6-dichloro-3-oxo-2,3-dihydro-1H-indene\\-2,1-diylidene))dimalononitrile
where X stands for four equivalent groups.  When we want to be more specific,
then we will write IDIC(X)-4Cl.  For example, X should be replaced by 
``tetrahexyl'' in the original IDIC-4Cl \cite{LMS+19}, by ``tetraoctyl'' by
the compound in Refs.~\cite{WDZ+20,ZWD+20,HZLZ+21}, by ``tetrakis(4-hexylphenyl)'' 
in the compound in Ref.~\cite{QCZ+18}, and by ``tetramethyl'' for the 
simplified model compound for which calculations are actually done in 
this paper.

\item[PM6]
poly[(2,6-(4,8-bis(5-(2-ethylhexyl-3-fluoro)thiophen-2-yl)-benzo[1,2-b:4,5-b']-dithiophene))-{\em alt}-(5,5-(1',3'-di-2-thienyl-5',7'-bis(2-ethylhexyl)benzo[1',2'-c:4',5'-c']dithiophene-4,8-dione)]
\begin{center}
\includegraphics[width=0.4\textwidth]{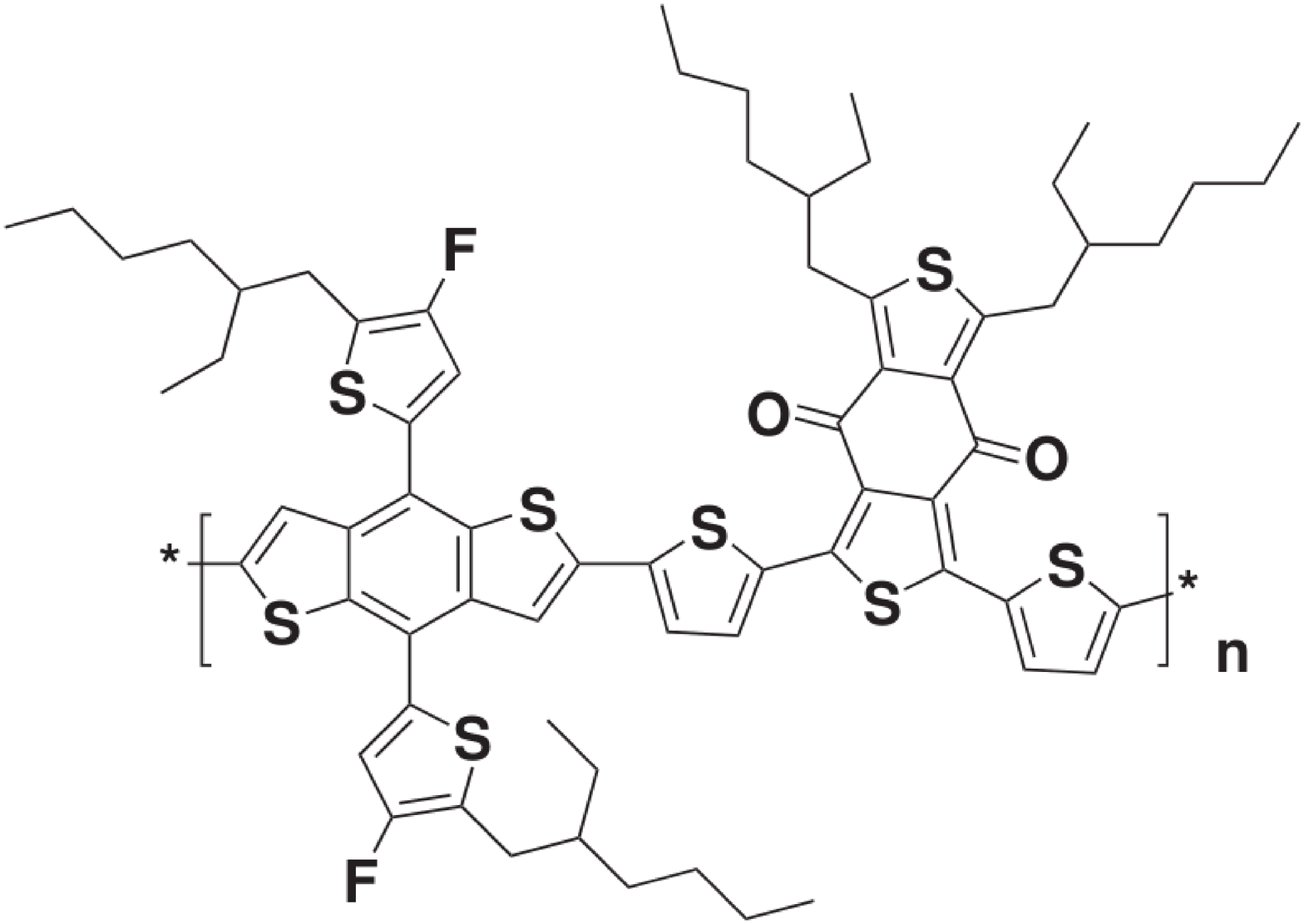}
\end{center}

\end{description}

\section{Optimized Structures}

In this appendix, we give the $(x,y,z)$-coordinates in 
{\AA}ngstr\"oms of our optimized geometries.  
First are the geometries optimized at the 
HSEH1PBE/6-311G(d,p)/PCM level.
Each geometry is followed by a table with the ten lowest vibrational 
frequencies {\em after} projecting out translations and rotations.

\subsection*{Optimized Geometries and Frequencies in Chloroform}

\begin{tabular}{cccc|cccc}
\hline \hline
\multicolumn{8}{c}{{\bf M$_0$} (simplified model of IDIC-4Cl)} \\
\multicolumn{8}{c}{Optimized Geometry in Chloroform} \\
Atom (No.) & $x$       & $y$      & $z$  &
Atom (No.) & $x$       & $y$      & $z$  \\
\hline
C(1)      & -5.057198 &  1.366294 & -0.018172 &
H(41)     & -1.347859 &  2.725303 & -2.238766 \\
S(2)      & -4.265293 & -0.209939 &  0.018629 &
H(42)     &  4.406826 & -3.440816 &  0.074379 \\
C(3)      & -2.735637 &  0.549772 & -0.006033 &
H(43)     &  1.328323 & -2.856013 & -2.089611 \\
C(4)      & -2.799539 &  1.938500 & -0.043460 &
H(44)     &  1.827425 & -4.282161 & -1.160868 \\
C(5)      & -4.104005 &  2.400601 & -0.049474 &
H(45)     &  0.129976 & -3.783035 & -1.167815 \\
\hline \hline
\end{tabular}

\begin{tabular}{cc|cc}
\hline \hline
\multicolumn{4}{c}{{\bf M$_0$} (simplified model of IDIC-4Cl)} \\
\multicolumn{4}{c}{Lowest Vibrational Frequencies in Chloroform} \\
Vibration Number & Frequency & 
Vibration Number & Frequency \\
\hline
1 &  4.150 cm$^{-1}$ &
6 & 23.890 cm$^{-1}$ \\
2 & 4.520 cm$^{-1}$ &
7 & 33.150 cm$^{-1}$ \\
3 & 6.650 cm$^{-1}$ &
8 & 37.380 cm$^{-1}$ \\
4 & 15.310 cm$^{-1}$ &
9 & 38.690 cm$^{-1}$ \\
5 & 17.700 cm$^{-1}$ &
10 & 39.200 cm$^{-1}$ \\
\hline \hline
\end{tabular}
\\ The largest magnitude imaginary frequency before projecting out the
vibrations and rotations was $8.9994i$ cm$^{-1}$.

\begin{tabular}{cccc|cccc}
\hline \hline
\multicolumn{8}{c}{{\bf M$_1$} (simplified model of IDIC-4CF$_3$)} \\
\multicolumn{8}{c}{Optimized Geometry in Chloroform} \\
Atom (No.) & $x$       & $y$      & $z$  &
Atom (No.) & $x$       & $y$      & $z$  \\
\hline
C(1)       &   7.520153  &  -1.335935  &  -0.057351 &
H(47)      &  -0.070975  &  -3.746423  &  -1.383606 \\
C(2)       &   8.851436  &  -1.916614  &  -0.044065 & 
H(48)      &   1.185252  &  -2.859633  &  -2.267388 \\
C(3)       &   7.661579  &   0.126403  &  -0.030656 &
C(49)      &  -1.359283  &   0.184278  &  -0.101061 \\
C(4)       &   9.817140  &  -0.793386  &   0.008643 &
H(50)      &  -1.559063  &  -1.971783  &  -0.100614 \\
C(5)       &   9.112923  &   0.406931  &   0.013181 &
C(51)      &  -1.269765  &   2.616637  &  -0.103351 \\
\hline \hline
\end{tabular}

\begin{tabular}{cc|cc}
\hline \hline
\multicolumn{4}{c}{{\bf M$_1$} (simplified model of IDIC-4CF$_3$)} \\
\multicolumn{4}{c}{Lowest Vibrational Frequencies in Chloroform} \\
Vibration Number & Frequency & 
Vibration Number & Frequency \\
\hline
1 & 4.440 cm$^{-1}$ &
6 & 18.980 cm$^{-1}$ \\
2 & 7.990 cm$^{-1}$ &
7 & 27.220 cm$^{-1}$ \\
3 & 8.130 cm$^{-1}$ &
8 & 27.590 cm$^{-1}$ \\
4 & 13.590 cm$^{-1}$ &
9 & 33.120 cm$^{-1}$ \\
5 & 14.760 cm$^{-1}$ &
10 & 35.880 cm$^{-1}$ \\
\hline \hline
\end{tabular}
\\ The largest magnitude imaginary frequency before projecting out the
vibrations and rotations was $3.7577i$ cm$^{-1}$.

\begin{tabular}{cccc|cccc}
\hline \hline
\multicolumn{8}{c}{{\bf M$_2$} (simplified model of IDIC-4(CH)$_2$)} \\
\multicolumn{8}{c}{Optimized Geometry in Chloroform} \\
Atom (No.) & $x$       & $y$      & $z$  &
Atom (No.) & $x$       & $y$      & $z$  \\
\hline 
C(1)  &   7.594014 &  -0.884922 &   0.056250 &
H(47) &   0.150433 &  -3.748331 &  -1.126997 \\
C(2)  &   8.958424 &  -1.390617 &   0.027047 &
H(48) &   1.342347 &  -2.789427 &  -2.024241 \\
C(3)  &   7.648718 &   0.586560 &   0.013275 &
C(49) &  -1.369337 &   0.099903 &   0.139879 \\
C(4)  &   9.857478 &  -0.225972 &  -0.073701 &
H(50) &  -1.432972 &  -2.064659 &   0.140141 \\
C(5)  &   9.069426 &   0.955783 &  -0.074097 &
C(51) &  -1.429705 &   2.532341 &   0.140758 \\
\hline \hline
\end{tabular}

\begin{tabular}{cc|cc}
\hline \hline
\multicolumn{4}{c}{{\bf M$_2$} (simplified model of IDIC-4(CH)$_2$)} \\
\multicolumn{4}{c}{Lowest Vibrational Frequencies in Chloroform} \\
Vibration Number & Frequency & 
Vibration Number & Frequency \\
\hline
1 & 1.430 cm$^{-1}$ &
6 & 25.490 cm$^{-1}$ \\
2 & 5.240 cm$^{-1}$ &
7 & 34.750 cm$^{-1}$ \\
3 & 10.230 cm$^{-1}$ &
8 & 38.020 cm$^{-1}$ \\
4 & 16.800 cm$^{-1}$ &
9 & 38.410 cm$^{-1}$ \\
5 & 18.650 cm$^{-1}$ &
10 & 45.700 cm$^{-1}$ \\
\hline \hline
\end{tabular}
\\ The largest magnitude imaginary frequency before projecting out the
vibrations and rotations was $10.6013i$ cm$^{-1}$.

\begin{tabular}{cccc|cccc}
\hline \hline
\multicolumn{8}{c}{{\bf M$_3$} (simplified model of IDIC-4CN)} \\
\multicolumn{8}{c}{Optimized Geometry in Chloroform} \\
Atom (No.) & $x$       & $y$      & $z$  &
Atom (No.) & $x$       & $y$      & $z$  \\
\hline
\hline \hline
C(1)  &  -7.593683 &   -0.816951 &   0.022905 & 
H(43) &  -0.183140 &   -3.742649 &   1.310845 \\
C(2)  &  -8.959586 &   -1.307587 &   0.026822 & 
H(44) &  -1.373846 &   -2.772864 &   2.197903 \\
C(3)  &  -7.634529 &    0.649003 &  -0.007058 & 
C(45) &   1.368563 &    0.090063 &   0.030230 \\
C(4)  &  -9.851551 &   -0.124238 &  -0.020659 & 
H(46) &   1.418925 &   -2.075279 &   0.030177 \\
C(5)  &  -9.064729 &    1.030428 &  -0.036964 &
C(47) &   1.446235 &    2.523410 &   0.032636 \\
\hline \hline
\end{tabular}

\begin{tabular}{cc|cc}
\hline \hline
\multicolumn{4}{c}{{\bf M$_3$} (simplified model of IDIC-4CN)} \\
\multicolumn{4}{c}{Lowest Vibrational Frequencies in Chloroform} \\
Vibration Number & Frequency & 
Vibration Number & Frequency \\
\hline
1 & 3.860 cm$^{-1}$ &
6 & 26.260 cm$^{-1}$ \\
2 & 4.160 cm$^{-1}$ &
7 & 33.930 cm$^{-1}$ \\
3 & 10.850 cm$^{-1}$ &
8 & 37.730 cm$^{-1}$ \\
4 & 17.440 cm$^{-1}$ &
9 & 38.180 cm$^{-1}$ \\
5 & 18.120 cm$^{-1}$ &
10 & 41.320 cm$^{-1}$ \\
\hline \hline
\end{tabular}
\\ The largest magnitude imaginary frequency before projecting out the
vibrations and rotations was $9.7885i$ cm$^{-1}$.

\begin{tabular}{cccc|cccc}
\hline \hline
\multicolumn{8}{c}{{\bf M$_4$} (simplified model of IDIC-2CN2COOCH3)} \\
\multicolumn{8}{c}{Optimized Geometry in Chloroform} \\
Atom (No.) & $x$       & $y$      & $z$  &
Atom (No.) & $x$       & $y$      & $z$  \\
\hline
\hline \hline
C(1)   &   7.519631 &   1.337113 &   0.034520 &
H(48)  &  -0.074437 &  3.746708  &   1.337696 \\
C(2)   &   8.851482 &   1.916545 &   0.030976 &
H(49)  &   1.181142 &  2.861152  &   2.223611 \\
C(3)   &   7.659274 &  -0.125118 &   0.009363 &
C(50)  &  -1.359196 & -0.184991  &   0.056642 \\
C(4)   &   9.817536 &   0.793237 &  -0.013413 &
H(51)  &  -1.559972 &  1.970989  &   0.056663 \\
C(5)   &   9.109046 &  -0.410421 &  -0.022903 &
C(52)  &  -1.268819 & -2.617347  &   0.058456 \\
\hline \hline
\end{tabular}

\begin{tabular}{cc|cc}
\hline \hline
\multicolumn{4}{c}{{\bf M$_4$} (simplified model of IDIC-2CN2COOCH3)} \\
\multicolumn{4}{c}{Lowest Vibrational Frequencies in Chloroform} \\
Vibration Number & Frequency & 
Vibration Number & Frequency \\
\hline
1 & 3.850 cm$^{-1}$ &
6 & 20.570 cm$^{-1}$ \\
2 & 7.770 cm$^{-1}$ &
7 & 27.970 cm$^{-1}$ \\
3 & 8.690 cm$^{-1}$ &
8 & 28.530 cm$^{-1}$ \\
4 & 14.820 cm$^{-1}$ &
9 & 32.350 cm$^{-1}$ \\
5 & 15.740 cm$^{-1}$ &
10 & 34.040 cm$^{-1}$ \\
\hline \hline
\end{tabular}
\\ The largest magnitude imaginary frequency before projecting out the
vibrations and rotations was $4.2376i$ cm$^{-1}$.

\subsection*{Optimized Geometries and Frequencies in Acetonitrile}

\begin{tabular}{cccc|cccc}
\hline \hline
\multicolumn{8}{c}{{\bf M$_0$} (simplified model of IDIC-4Cl)} \\
\multicolumn{8}{c}{Optimized Geometry in Acetonitrile} \\
Atom (No.) & $x$       & $y$      & $z$  &
Atom (No.) & $x$       & $y$      & $z$  \\
\hline
\hline \hline
C(1)   &  -7.585660 &   -0.919075 &    0.014336 &
H(41)  &  -0.131945 &   -3.745730 &    1.258926 \\
C(2)   &  -8.946001 &   -1.427661 &    0.034197 &
H(42)  &  -1.332086 &   -2.791811 &    2.151026 \\
C(3)   &  -7.650889 &    0.549840 &   -0.012316 &
C(43)  &   1.367603 &    0.107462 &   -0.016583 \\
C(4)   &  -9.851922 &   -0.258608 &    0.006793 &
H(44)  &   1.445293 &   -2.056938 &   -0.016564 \\
C(5)   &  -9.082570 &    0.907616 &   -0.018354 &
C(45)  &   1.414363 &    2.541193 &   -0.014200 \\
\hline \hline
\end{tabular}

\begin{tabular}{cc|cc}
\hline \hline
\multicolumn{4}{c}{{\bf M$_0$} (simplified model of IDIC-4Cl)} \\
\multicolumn{4}{c}{Lowest Vibrational Frequencies in Acetonitrile} \\
Vibration Number & Frequency & 
Vibration Number & Frequency \\
\hline
1 & 4.720 cm$^{-1}$ &
6 & 23.500 cm$^{-1}$ \\
2 & 8.280 cm$^{-1}$ &
7 & 33.200 cm$^{-1}$ \\
3 & 9.790 cm$^{-1}$ &
8 & 37.870 cm$^{-1}$ \\
4 & 16.010 cm$^{-1}$ &
9 & 38.030 cm$^{-1}$ \\
5 & 17.760 cm$^{-1}$ &
10 & 38.590 cm$^{-1}$ \\
\hline \hline
\end{tabular}
\\ The largest magnitude imaginary frequency before projecting out the
vibrations and rotations was $6.9659i$ cm$^{-1}$.

\begin{tabular}{cccc|cccc}
\hline \hline
\multicolumn{8}{c}{{\bf M$_1$} (simplified model of IDIC-4CF$_3$)} \\
\multicolumn{8}{c}{Optimized Geometry in Acetonitrile} \\
Atom (No.) & $x$       & $y$      & $z$  &
Atom (No.) & $x$       & $y$      & $z$  \\
\hline
C(1)    &   7.520209 &  -1.330429 &  -0.067698 &
H(47)   &  -0.069776 &  -3.745958 &  -1.395265 \\
C(2)    &   8.849228 &  -1.911726 &  -0.054862 &
H(48)   &   1.186924 &  -2.860078 &  -2.279811 \\
C(3)    &   7.662748 &   0.131046 &  -0.034497 &
C(49)   &  -1.358913 &   0.183454 &  -0.113507 \\
C(4)    &   9.816772 &  -0.790399 &   0.009187 &
H(50)   &  -1.558395 &  -1.972884 &  -0.113973 \\
C(5)    &   9.114431 &   0.410635 &   0.016622 &
C(51)   &  -1.270630 &   2.616419 &  -0.114195 \\
\hline \hline
\end{tabular}

\begin{tabular}{cc|cc}
\hline \hline
\multicolumn{4}{c}{{\bf M$_1$} (simplified model of IDIC-4CF$_3$)} \\
\multicolumn{4}{c}{Lowest Vibrational Frequencies in Acetonitrile} \\
Vibration Number & Frequency & 
Vibration Number & Frequency \\
\hline
1 & 4.590 cm$^{-1}$ &
6 & 18.960 cm$^{-1}$ \\
2 & 7.960 cm$^{-1}$ &
7 & 27.230 cm$^{-1}$ \\
3 & 9.020 cm$^{-1}$ &
8 & 27.820 cm$^{-1}$ \\
4 & 13.820 cm$^{-1}$ &
9 & 33.460 cm$^{-1}$ \\
5 & 14.770 cm$^{-1}$ &
10 & 35.740 cm$^{-1}$ \\
\hline \hline
\end{tabular}
\\ The largest magnitude imaginary frequency before projecting out the
vibrations and rotations was $3.4736i$ cm$^{-1}$.

\begin{tabular}{cccc|cccc}
\hline \hline
\multicolumn{8}{c}{{\bf M$_2$} (simplified model of IDIC-4(CH)$_2$)} \\
\multicolumn{8}{c}{Optimized Geometry in Acetonitrile} \\
Atom (No.) & $x$       & $y$      & $z$  &
Atom (No.) & $x$       & $y$      & $z$  \\
\hline 
C(1)  &   -7.592522 &   0.881024 &   0.030538 &
H(47) &   -0.128274 &   3.819404 &  -0.906000 \\
C(2)  &   -8.955193 &   1.387205 &  -0.002070 &
H(48) &   -1.308934 &   2.911886 &  -1.876625 \\
C(3)  &   -7.648312 &  -0.589684 &  -0.038215 &
C(49) &    1.369991 &  -0.109085 &   0.165524 \\
C(4)  &   -9.855196 &   0.224757 &  -0.096296 &
H(50) &    1.433619 &   2.055684 &   0.278750 \\
C(5)  &   -9.068355 &  -0.957957 &  -0.106680 &
C(51) &    1.431295 &  -2.541967 &   0.041303 \\
\hline \hline
\end{tabular}

\begin{tabular}{cc|cc}
\hline \hline
\multicolumn{4}{c}{{\bf M$_2$} (simplified model of IDIC-4(CH)$_2$)} \\
\multicolumn{4}{c}{Lowest Vibrational Frequencies in Acetonitrile} \\
Vibration Number & Frequency & 
Vibration Number & Frequency \\
\hline
1 & 4.020 cm$^{-1}$ &
6 & 27.750 cm$^{-1}$ \\
2 & 7.190 cm$^{-1}$ &
7 & 34.570 cm$^{-1}$ \\
3 & 17.070 cm$^{-1}$ &
8 & 35.410 cm$^{-1}$ \\
4 & 17.800 cm$^{-1}$ &
9 & 40.040 cm$^{-1}$ \\
5 & 19.750 cm$^{-1}$ &
10 & 47.860 cm$^{-1}$ \\
\hline \hline
\end{tabular}
\\ The largest magnitude imaginary frequency before projecting out the
vibrations and rotations was $5.4839i$ cm$^{-1}$.

\begin{tabular}{cccc|cccc}
\hline \hline
\multicolumn{8}{c}{{\bf M$_3$} (simplified model of IDIC-4CN)} \\
\multicolumn{8}{c}{Optimized Geometry in Acetonitrile} \\
Atom (No.) & $x$       & $y$      & $z$  &
Atom (No.) & $x$       & $y$      & $z$  \\
\hline
\hline \hline
C(1)  &  -7.593011 &  -0.812876 &   0.027051 &
H(43) &  -0.183863 &  -3.742581 &   1.317509 \\
C(2)  &  -8.956631 &  -1.304164 &   0.030424 &
H(44) &  -1.375106 &  -2.773393 &   2.205039 \\
C(3)  &  -7.634848 &   0.652298 &  -0.007674 &
C(45) &   1.368068 &   0.089407 &   0.038217 \\
C(4)  &  -9.850535 &  -0.122359 &  -0.025936 &
H(46) &   1.418540 &  -2.076224 &   0.038008 \\
C(5)  &  -9.065599 &   1.033042 &  -0.043988 &
C(47) &   1.446355 &   2.523288 &   0.040053 \\
\hline \hline
\end{tabular}

\begin{tabular}{cc|cc}
\hline \hline
\multicolumn{4}{c}{{\bf M$_3$} (simplified model of IDIC-4CN)} \\
\multicolumn{4}{c}{Lowest Vibrational Frequencies in Acetonitrile} \\
Vibration Number & Frequency & 
Vibration Number & Frequency \\
\hline
1 &  3.620 cm$^{-1}$ &
6 & 26.320 cm$^{-1}$ \\
2 &  4.360 cm$^{-1}$ &
7 & 34.040 cm$^{-1}$ \\
3 & 12.600 cm$^{-1}$ &
8 & 37.380 cm$^{-1}$ \\
4 & 18.140 cm$^{-1}$ &
9 & 37.670 cm$^{-1}$ \\
5 & 18.380 cm$^{-1}$ &
10 & 41.820 cm$^{-1}$ \\
\hline \hline
\end{tabular}
\\ The largest magnitude imaginary frequency before projecting out the
vibrations and rotations was $10.4223i$ cm$^{-1}$.

\begin{tabular}{cccc|cccc}
\hline \hline
\multicolumn{8}{c}{{\bf M$_4$} (simplified model of IDIC-2CN2COOCH3)} \\
\multicolumn{8}{c}{Optimized Geometry in Acetonitrile} \\
Atom (No.) & $x$       & $y$      & $z$  &
Atom (No.) & $x$       & $y$      & $z$  \\
\hline
\hline \hline
C(1)   &   7.518901 &   1.332940 &   0.044240 &
H(48)  &  -0.073853 &   3.746005 &   1.355334 \\
C(2)   &   8.848337 &   1.912808 &   0.039040 &
H(49)  &   1.182791 &   2.861290 &   2.241146 \\
C(3)   &   7.659352 &  -0.128301 &   0.011772 &
C(50)  &  -1.358617 &  -0.184475 &   0.074289 \\
C(4)   &   9.816149 &   0.791310 &  -0.018627 &
H(51)  &  -1.559974 &   1.971731 &   0.074760 \\
C(5)   &   9.109629 &  -0.413013 &  -0.030413 &
C(52)  &  -1.268685 &  -2.617488 &   0.075074 \\
\hline \hline
\end{tabular}

\begin{tabular}{cc|cc}
\hline \hline
\multicolumn{4}{c}{{\bf M$_4$} (simplified model of IDIC-2CN2COOCH3)} \\
\multicolumn{4}{c}{Lowest Vibrational Frequencies in Acetonitrile} \\
Vibration Number & Frequency & 
Vibration Number & Frequency \\
\hline
1 & 4.370 cm$^{-1}$ &
6 & 20.940 cm$^{-1}$ \\
2 & 7.880 cm$^{-1}$ &
7 & 28.480 cm$^{-1}$ \\
3 & 9.880 cm$^{-1}$ &
8 & 28.620 cm$^{-1}$ \\
4 & 15.300 cm$^{-1}$ &
10 & 35.650 cm$^{-1}$ \\
5 & 15.680 cm$^{-1}$ &
9 & 33.300 cm$^{-1}$ \\
\hline \hline
\end{tabular}
\\ The largest magnitude imaginary frequency before projecting out the
vibrations and rotations was $3.2520i$ cm$^{-1}$.

\section{FMO theory diagram}
\begin{figure}[ht]
\begin{center}
\includegraphics[width=0.6\textwidth]{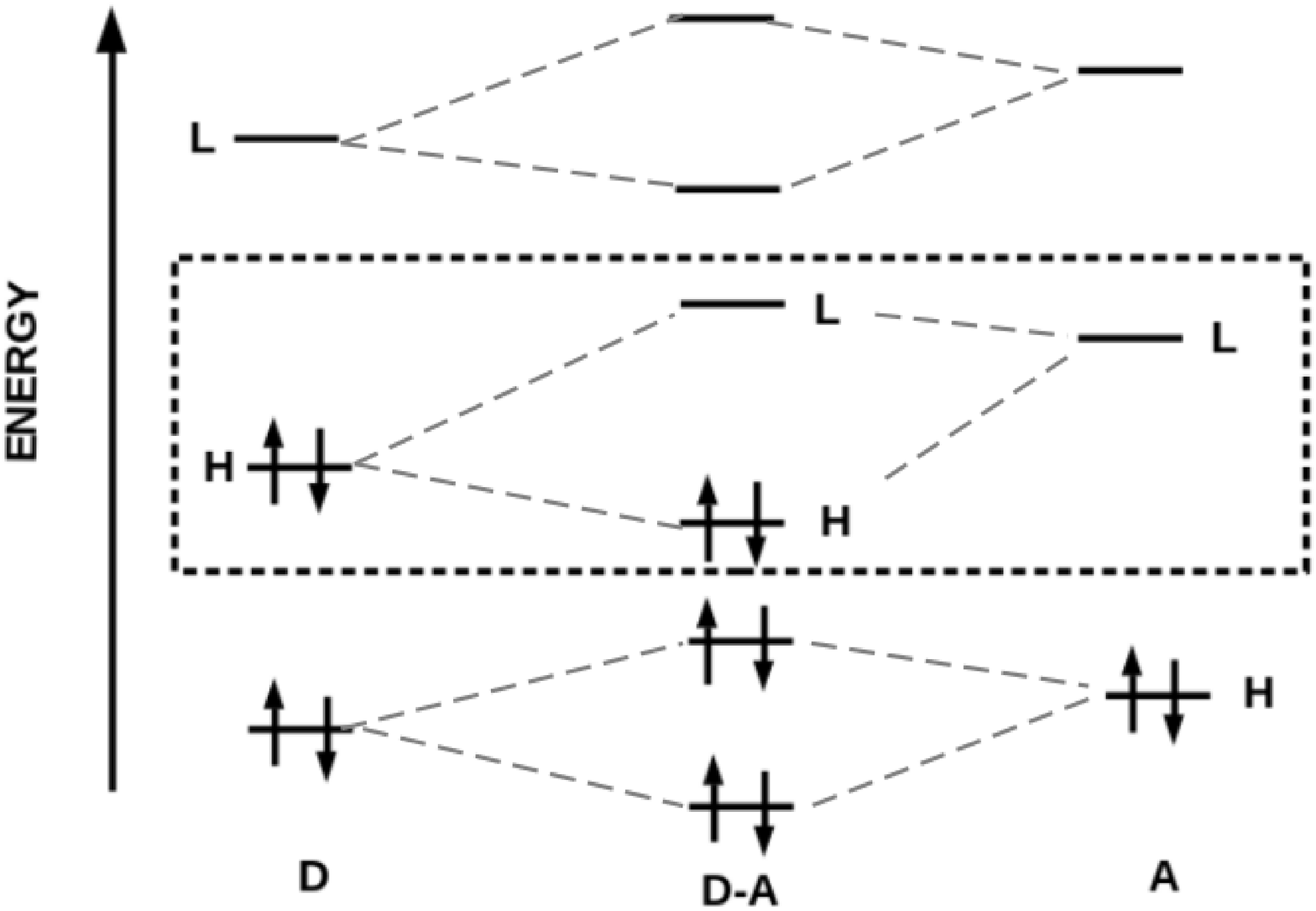}
\end{center}
\caption{
Generic FMO theory diagram \cite{F76,A07} with donor D and acceptor A 
joining to form a new D-A molecule.  H indicates the HOMO level and L
indicates the LUMO level.  Adding an acceptor to D decreases the LUMO
energy of D-A more than it decreases the HOMO energy.  Adding a donor to
A increases the HOMO energy of D-A more than it increases the LUMO
energy.
\label{fig:FMOtheory}
}
\end{figure}
The electronegativity of chlorine makes chloro groups electron acceptors
by induction (-I, $\sigma$-acceptor).  The presence of $p$-electrons 
\marginpar{\color{blue} $\pm$I, $\pm$M}
may also make chloro groups electron donors by resonance (+M, 
$\pi$-donor).  Going from IDIC to IDIC-4Cl lowers both the CV HOMO 
and the CV LUMO energies.  According to FMO theory
({\bf Fig.~\ref{fig:FMOtheory}}), this means that the +M effect is
more important than the -I effect as is usually the case.
FMO theory also seems to suggest that the CV HOMO-LUMO gap should
decrease on adding either an electron acceptor or an electron donor.
In this case, we see that adding chloro groups has decreased the 
CV HOMO-LUMO gap.

\section{Spectral Shift}

\begin{center}
\begin{tabular}{ccccc}
\hline \hline
X & $\lambda_{\mbox{max}}^{\mbox{soln}}$ & $\lambda_{\mbox{max}}^{\mbox{film}}$ & 
red shift & Reference \\
\hline 
\multicolumn{5}{c}{IDIC(X)}\\
tetrakis(4-hexylphenyl) & 647 nm   & 682 nm   & 0.098 eV & \cite{QCZ+18} \\
tetraoctyl              & 660 nm   & 710 nm   & 0.156 eV & \cite{HZLZ+21} \\
tetrahexyl              & 670 nm   & 720 nm   & 0.129 eV & \cite{LHZ+16} \\
\multicolumn{5}{c}{IDIC(X)-4Cl}\\
tetrakis(4-hexylphenyl) & 669 nm   & 709 nm   & 0.105 eV & \cite{QCZ+18} \\
tetraoctyl              & 700 nm   & 760 nm   & 0.140 eV & \cite{WDZ+20} \\
tetraoctyl              & 690 nm   & 760 nm   & 0.165 eV & \cite{ZWD+20} \\
tetraoctyl              & 690 nm   & 760 nm   & 0.165 eV & \cite{HZLZ+21} \\
tetrahexyl              & 690 nm   & 760 nm   & 0.165 eV & \cite{LMS+19} \\
\hline \hline
\end{tabular}
\end{center}

This table shows the red shift of the main peak in the absorption spectra in going
from chloroform solution to thin film. The X in IDIC(X) and in IDIC(X)-4Cl is chosen to control aggregation
which is one of the many parameters which may be important in organic
solar cells.  This aggregation is even more important when thin films
are formed. A larger energy shift 
is expected to correspond to closer approach of the planar polycyclic
core of the molecules.  Interestingly, but not unexpectedly, this shift 
is smallest when X is largest --- that is for 
IDIC(tetrakis(4-hexylphenyl)) and for
IDIC(tetrakis(4-hexylphenyl))-4Cl.

\singlespace
